# AI Sovereignty: A Qualitative Model of Strategic Competition as AI Becomes an Instrument of National Power

**Timothy Clancy *[a], Asmeret Naugle[b]**

[a] *Applied Research Laboratory for Intelligence and Security (ARLIS), University of Maryland, College Park, USA*

[b] *Sandia National Laboratories, Albuquerque, New Mexico, USA*

*Corresponding Author: Timothy Clancy timc@umd.edu

## Abstract

AI sovereignty is the extent to which a nation independently controls its artificial intelligence (AI) technologies. The race toward ever-more-sophisticated frontier AI models is of increasing strategic importance, with nations considering how AI might improve their economic situations, competitive advantage, and overall national power. However, the costs of AI sovereignty are enormous, and we lack definitions and conceptual models to navigate evolving AI sovereignty dynamics. We address this gap with definitions relevant to AI sovereignty, along with a first-of-its-kind qualitative model that incorporates micro, meso, and macro contributors. Model-based qualitative forecasts highlight competitive dynamics and evolving potential for AI-driven national power. The model identifies key leverage points that nations can use to enhance their own growth or degrade an adversary's, including consideration of accelerators, electricity, water, data sets and skilled workforce. These leverage points can be activated at strategic and operational levels through both direct kinetic actions, such as Iran's targeting of data centers with drones, and indirect non-kinetic effects including cyber, space, information, economic coercion and diplomacy. If our assumptions and hypotheses are valid, this strategic competition may come to define how nations improve their economic situations, competitive advantage, and overall national power in the 21st Century.

# Introduction

AI sovereignty is the extent to which a nation independently controls its artificial intelligence (AI) technologies [1, p. 14], encompassing the nation's capabilities in data, workforce, natural resources, infrastructure, and model training and hosting [1]. Recent advancements in AI technology have been driven by activities primarily concentrated in a few nations. However, as AI capabilities increase and become more established, additional nations will likely desire AI sovereignty and take actions to claim it.

Nations might strive for AI sovereignty because they expect it will improve their economic situations, competitive advantage, and overall national power [2]. National power encompasses all resources and capabilities a nation can access to pursue its goals [1], including diplomatic, informational, military, and economic instruments[3, p. 121].

Our hypothesis in this work builds from three key assumptions. First, we subsume all previous variations of AI into a broadly defined capability called ‘agentic AI’. Second, across successive generations of improvement agentic AI becomes an instrument of national power leading to strategic competition. Third, that competition may result in the use of diplomatic, economic, and military measures [4, p. 63] to manipulate access to AI across individual nations, regional hegemons, and alliance-based spheres of influence.

Under these assumptions there is a national security need to define, measure, and forecast AI sovereignty and its impact on national power. While early AI development has been heavily market-based, access to data center-generated compute and AI capabilities may move toward a future shaped by strategic interest, potentially producing more fragmented or asymmetrical structures. Nations, and alliances, with relatively more AI sovereignty will hold the strategic advantage over those without, with non-AI-sovereign nations risking disadvantage and accepting foreign supply chain risks.

This paper presents a qualitative system dynamics model designed as an initial step toward untangling the key dynamics of AI sovereignty and its effects on national power and strategic competition. We include definitions, metrics, and behavior modes relevant to the investigation of these dynamics. We begin with a focus on within-nation dynamics and then extend the model to consider dynamics between two or more nation-states involved in strategic competition. Our focus is on national security, but the consideration of national power presented here is also useful for assessing economic or other competition or collaboration. This work is intended as an initial investigation and is not designed to exhaustively analyze the technical details of AI sovereignty.

We begin with a brief review of recent developments in AI capabilities, from generative to agentic AI. Next, we articulate micro, meso, and macro measures necessary to model the 'operational physics’ of AI Sovereignty, focused on the resources required for nations to develop AI capabilities. This includes associated costs in power, water, emissions, land, and spending. We then identify relationships between these variables in a qualitative system model using a causal loop diagram (CLD), paying special attention to competitive dynamics between a nation-state and its strategic adversary. The qualitative model is accompanied by reference modes describing key past and future behaviors. We conclude with a description of next steps toward converting the qualitative system model into a dynamic system dynamics simulation model.

## From Generative AI to Agentic AI

The latest revolution in AI began in late 2022 and has since progressed rapidly across many stages of capability. First, Generative AI (GenAI) used Large Language Models (LLMs) trained on massive-scale datasets to generate content in ways never seen before. GenAI extended into multimodal capabilities, able to handle diverse types of data. This phase of GenAI generated responses to each prompt. Still, it did not retain a longer memory or contemplation of that prompt. GenAI evolved into Reasoning AI capabilities capable of managing complex multi-stage efforts to interpret a prompt, design and evaluate plans of action to satisfy that prompt [5]. Tool-augmented AI Agents developed in parallel with Reasoning AI and incorporated capabilities for accessing and incorporating external tools. This allows for more current information, software execution, and dynamic interaction [5]. Hybrid Generative Agents combine the prompt-based content generation of GenAI with the individual autonomy of Tool-augmented AI Agents [5]. All of these prior capabilities have been incorporated into Agentic AI, autonomous AI systems capable of pursuing complex goals [6].

## Micro, Meso, and Macro Measures of AI Sovereignty

First, we consider AI sovereignty from a perspective of 'operational causality' [7]. Our model needs to, within reason, approximate the resources and mechanisms needed to develop and maintain AI capabilities. To achieve this, we define and qualitatively model these mechanisms at three scales: micro, meso, and macro. We take inspiration from population modeling [8], where single individuals or small groups (micro) form into larger communities or distinctive identities (meso) that, taken together, form an aggregate measure of a nation or region (macro). Figure 1 provides an overview of our conceptual approach to understanding the operational causality of micro-, meso-, and macro-level means of developing AI sovereignty and national power in Agentic AI.

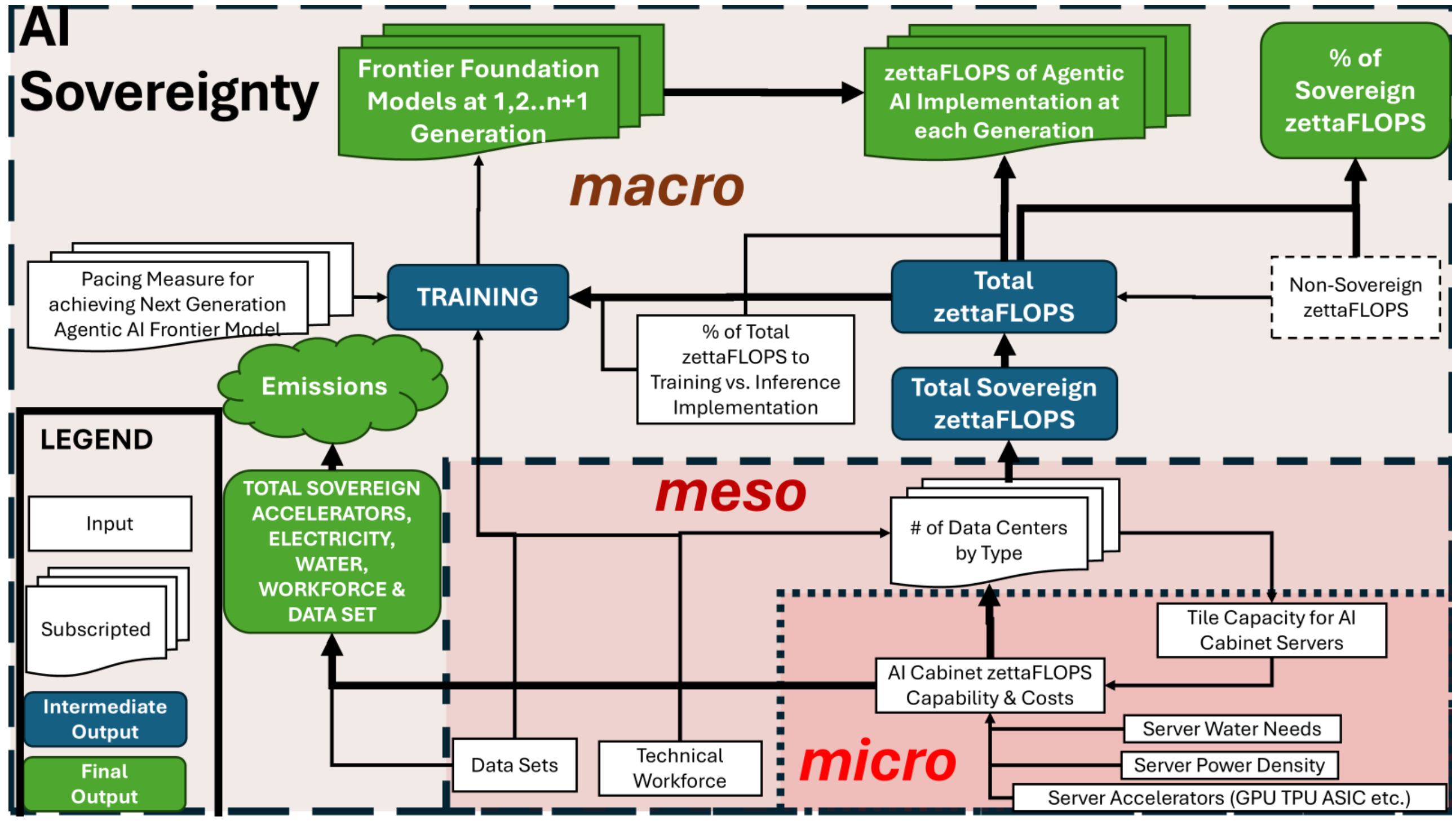

*Figure 1: Mechanisms determining AI sovereignty at the micro, meso, and macro levels*

At the micro and meso levels, we select key units of measure that represent both physical constraints and lasting containers. They are less likely to change, even as rapid innovation changes everything around them and through the container. At the micro level, the container is the AI Cabinet, a physical space with defined dimensions within which accelerators, electricity, water usage, and compute power can be calculated per cabinet. At the meso level, this unit of measure is the AI data center, which collects and houses AI Cabinets. Micro and meso for agentic AI Cabinets and Data Centers can likely be held constant over time, even as the volume of compute, as well as costs in dollars, electricity, water, and equipment, adjust dynamically. At the macro level, we disaggregate these physical constraints and use aggregate totals. Total compute power might be used to train frontier models and to implement AI capabilities.

We specify agentic AI Cabinets and agentic AI Data Centers to distinguish this new breed from past AI iterations. The advancement from previous versions of AI to agentic AI doesn't just increase potential capability – it also significantly changes the resources required and the constraints imposed. This is discussed in more detail in the supplementary materials. In summary, agentic AI requires much more electricity and generates much more heat, so AI Cabinets now require liquid cooling. And previous server cabinets and data centers may not have been designed for the high electrical loads or liquid cooling infrastructure. The increase in accelerators, electricity, and water – and the way these are delivered – means agentic AI may require new builds or significant upgrades to historical data centers before they are usable, so our measures of AI sovereignty and national power begin with agentic AI.

Since no model can perfectly represent the real world [9], our three most important boundary assumptions are the ones previously mentioned in the introduction. We recognize all three of these boundary assumptions are subject to extensive, and valid debate, but are outside the scope of this work. Additional factors outside our scope include advances in lithography for chip design, improvements in training and configuration methods for frontier models that may make them more efficient, and changes in data center design [10, pp. 3–5]. While future iterations may consider these potential advancements, we strive for a more compact model in terms of size, speed, and complexity. Likewise, degradation measures such as poisoned training datasets [11], compromised accelerator supply chains or other strategic countermeasures might be incorporated into future iterations or represented by aggregate modifiers.

Readers interested in a fuller treatment of the specific micro, meso, and macro phenomena modeled, including definitions, possible parameter ranges, and differences between agentic AI and previous AI in terms of resources, are directed to the supplementary materials.

# Proposed Qualitative Model of AI Sovereignty and National Power

We use a modified version of the system dynamics standard method to construct our qualitative model of AI sovereignty & national power in Agentic AI [25]. We incorporated the micro, meso, and macro measures of AI sovereignty discussed in the previous section into a qualitative model and then extended it to investigate the causal mechanisms underlying nations' AI sovereignty decision-making and the relevant interactions between nations. Figure 2 shows the causal loop diagram implementation of the measures discussed in the previous section. Phenomena contributing to micro measures are shown in black, those contributing to meso measures in purple, those contributing to macro measures in blue, and resource needs in green. The two key measures we are studying, AI sovereignty and national power in Agentic AI, are bolded and underlined. We assume that AI sovereignty and national power are determined by the focus nation's total zettaFLOPS of capability.

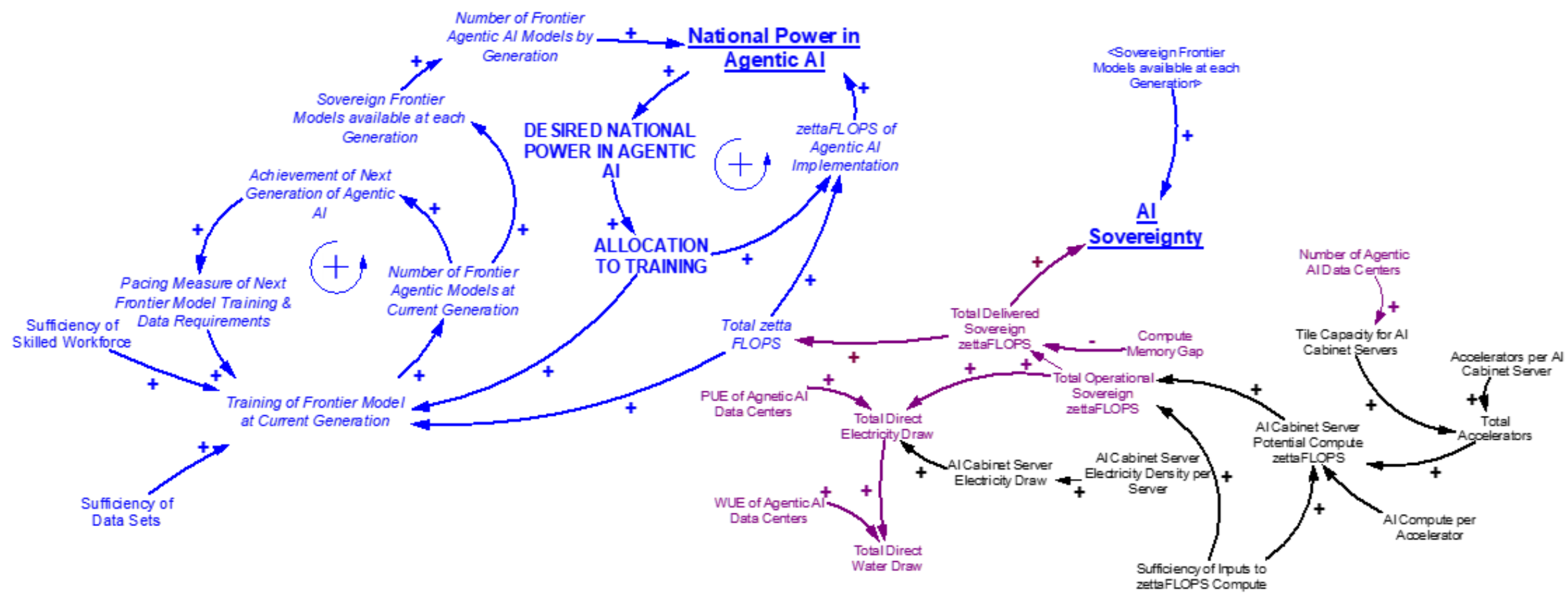

*Figure 2: Micro (black), meso (purple), and macro (blue) phenomena leading to national power and Sovereignty Measures*

Figure 3 considers mechanisms that limit the growth of AI capabilities, shown in black and underlined. A skilled workforce is necessary to develop and maintain AI capabilities. It can be created (with a delay) but at high cost. Electricity, water, and AI infrastructure are also significant needs. The amount of these resources made available for AI might be increased through sovereign efforts (blue capitalized). Still, tradeoffs with other potential uses must be considered. While these tradeoffs are not explicitly considered in this qualitative model, they may be explored in future iterations.

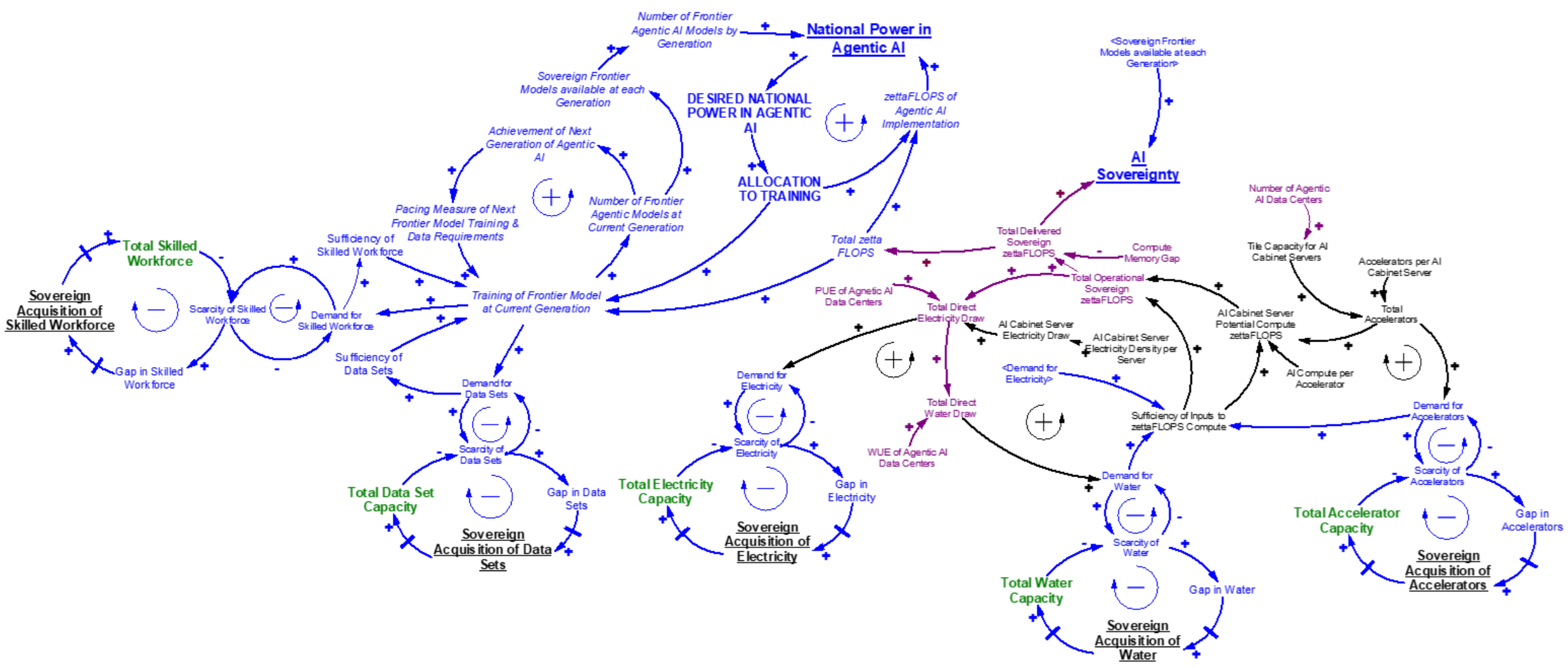

*Figure 3: Capacity Investment & Growth structures added with limits (black underlined), total capacities (green bold)*

In Figure 3 a series of generic structures representing the necessary investment to sufficiently generate inputs needed are added. These inputs include accelerators, water, and electricity for compute generation and skilled workforce and datasets for new frontier models. These generic structures are in the form of a growth and underinvestment structure, zoomed in on accelerators and water below in Figure 4.

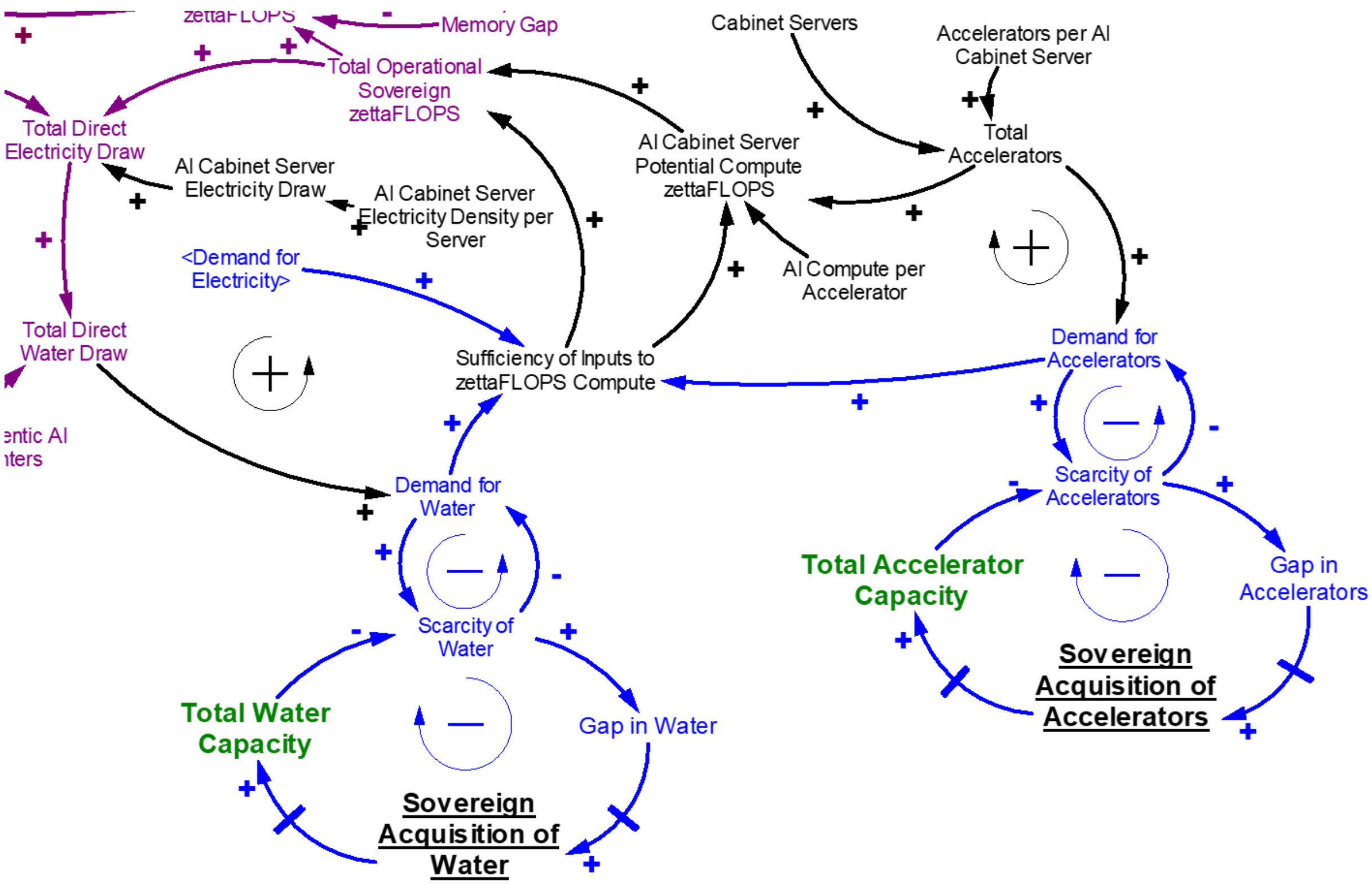

*Figure 4: Zoom in on growth and underinvestment structure of water and accelerators.*

A growth and underinvestment structure is a well-studied generic system archetype for behavior that is driven off of growth performance, in the case of Figure 4 the total accelerators and water draw respectively. This demand then identifies a scarcity relative to current capacity, creating a gap which increases acquisition and investment efforts. Failure to address the scarcity with an increase in capacity can destroy demand, creating a limit to growth, while meeting it generates more demand leading to more scarcity to be addressed in the futrue. The time delays between the identification of the gaps and additions in capacity; and the differences between time delays of the various types of inputs, create significant differences between how a sovereign and strategic adversary might decide to pull levers to achieve growth. Strength in accelerators may allow growth for a time but reach a limit in the deployment of water or electricity sufficient to utilize those accelerators in data centers.

Figure 5 adds consideration of foreign capabilities (in brown) and foreign competition (in red). The focus nation might leverage foreign-developed foundation models and infrastructure and recruit skilled AI labor from abroad. However, it will also need to consider the potential for competition with strategic adversaries. If the relative AI power of strategic adversaries becomes a focus of decision-making, AI arms races might ensue.

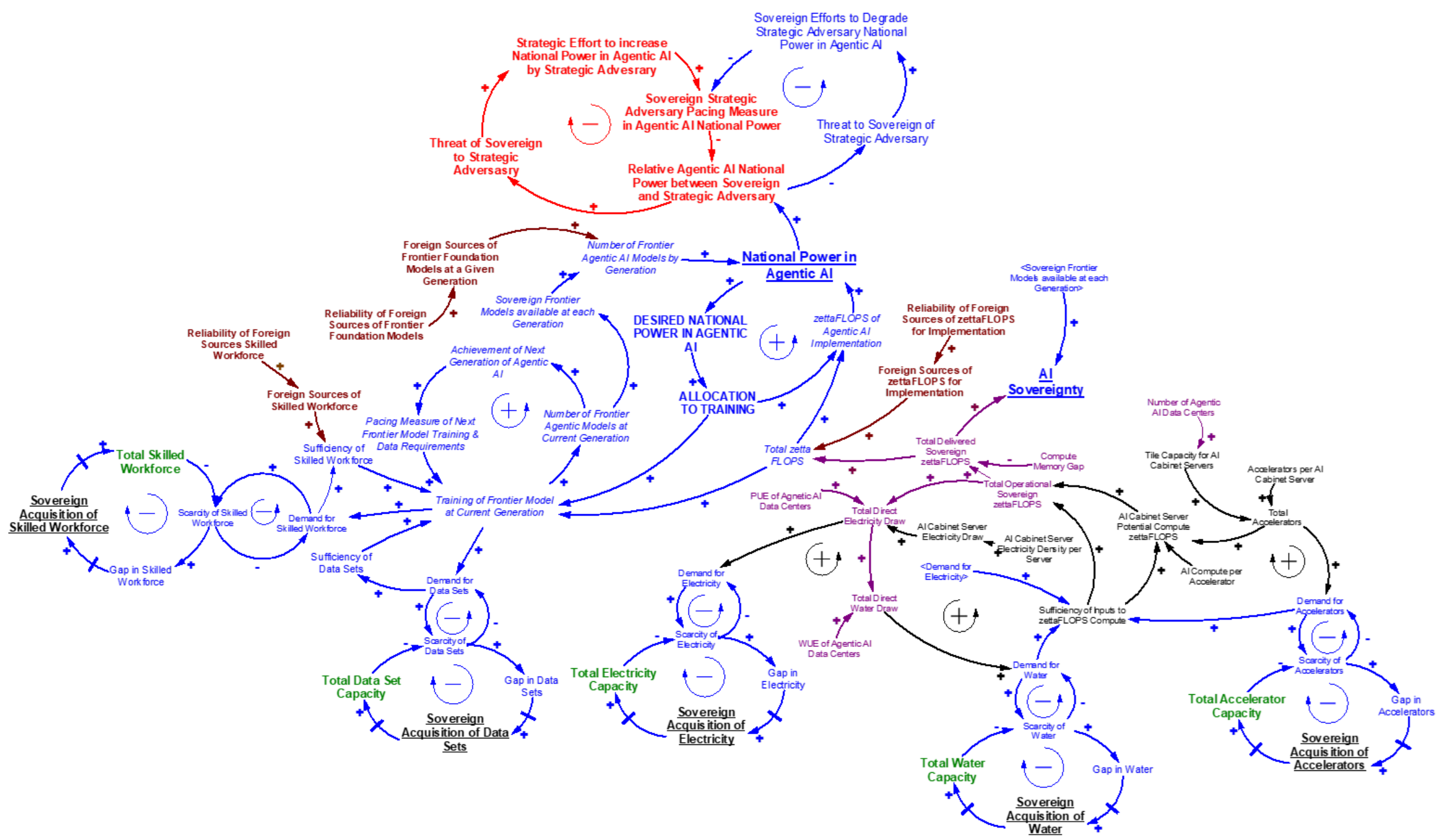

*Figure 5: Foreign capabilities (brown) and competition dynamics (red)*

Finally, Figure 6 introduces a strategic competition dynamic. The threat posed by an adversary may prompt a sovereign to adopt a strategy to pull some or all of the levers introduced in the focus nation's strategic efforts to increase or remove limits to AI growth. With a strong understanding of the limits to AI growth, the focus nation can allocate workforce, water, electricity, and infrastructure to AI efforts. However, if an AI arms race ignites between the focus nation and its strategic competitor, the competitor may be incentivized to attempt to prevent the focus nation from improving the situation, even within those limits.

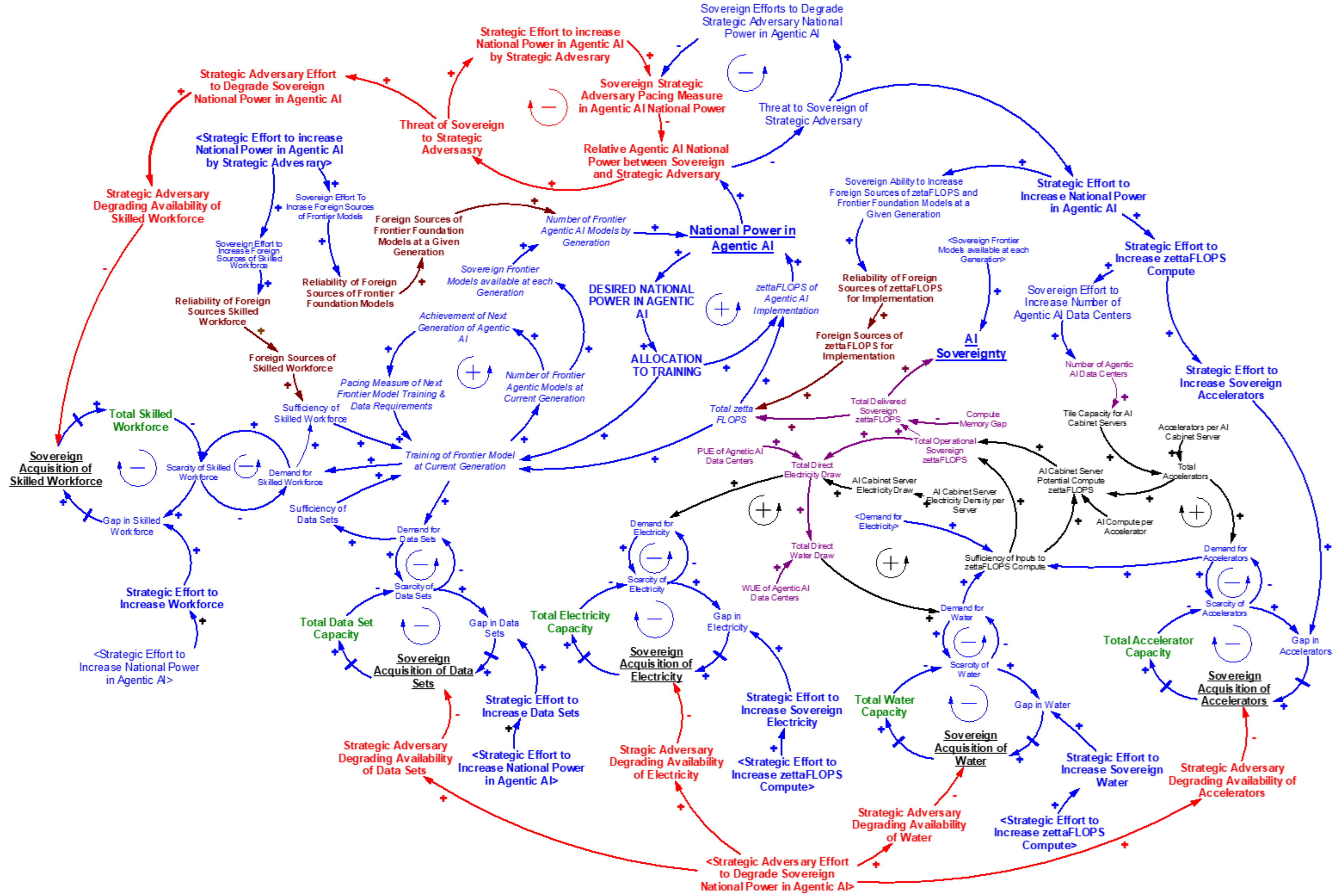

*Figure 6: Focus nation efforts to increase limits to growth and strategic adversary efforts to degrade them*

The strategic levers in Figure 6, for a focus nation to increase its resources while an adversary attempts to degrade them, represent the means of strategic competition. The levers are linked to micro and meso phenomena, which can be targeted as a means to the end of adjusting macro measures of AI sovereignty or national power in agentic AI. Although not presented for clarity, the model is symmetrical between the focus nation and the adversary, with each nation attempting to develop macro-national power in AI through its micro and meso means. This conceptualizes the full range of competition to increase a focus's capabilities while choosing which of an adversary's capabilities to degrade, while the adversary evaluates similar choices and selects its strategy. Which levers each nation pulls and why are determined in part by their natural advantages (e.g., cheaper and more plentiful electricity or access to accelerator manufacturing centers) as well as their national strategic policy and can be explored in future work.

# Hypothesized Dynamics (Reference Modes)

We have investigated relevant data and abstracted notional behavior modes that represent hypothesized dynamics of the system described in the causal loop diagrams above. Figure 7 shows hypothesized behavior over time of the growth of AI resources, including skilled labor, water, electricity, and infrastructure. System states of transition are presented as shaded horizontal bands on the charts. The lowest band indicates when sovereign resources are plentiful, while the second-lowest band indicates when internal limits are in place. The third band represents what can be obtained from foreign sources. The top band represents the resources needed for continued growth that are beyond sovereign and foreign limits. We presume s-shaped growth of these resources, with initial growth from within the focus nation leveling off when limits are reached. However, if foreign resources are leveraged, growth might continue longer before leveling off.

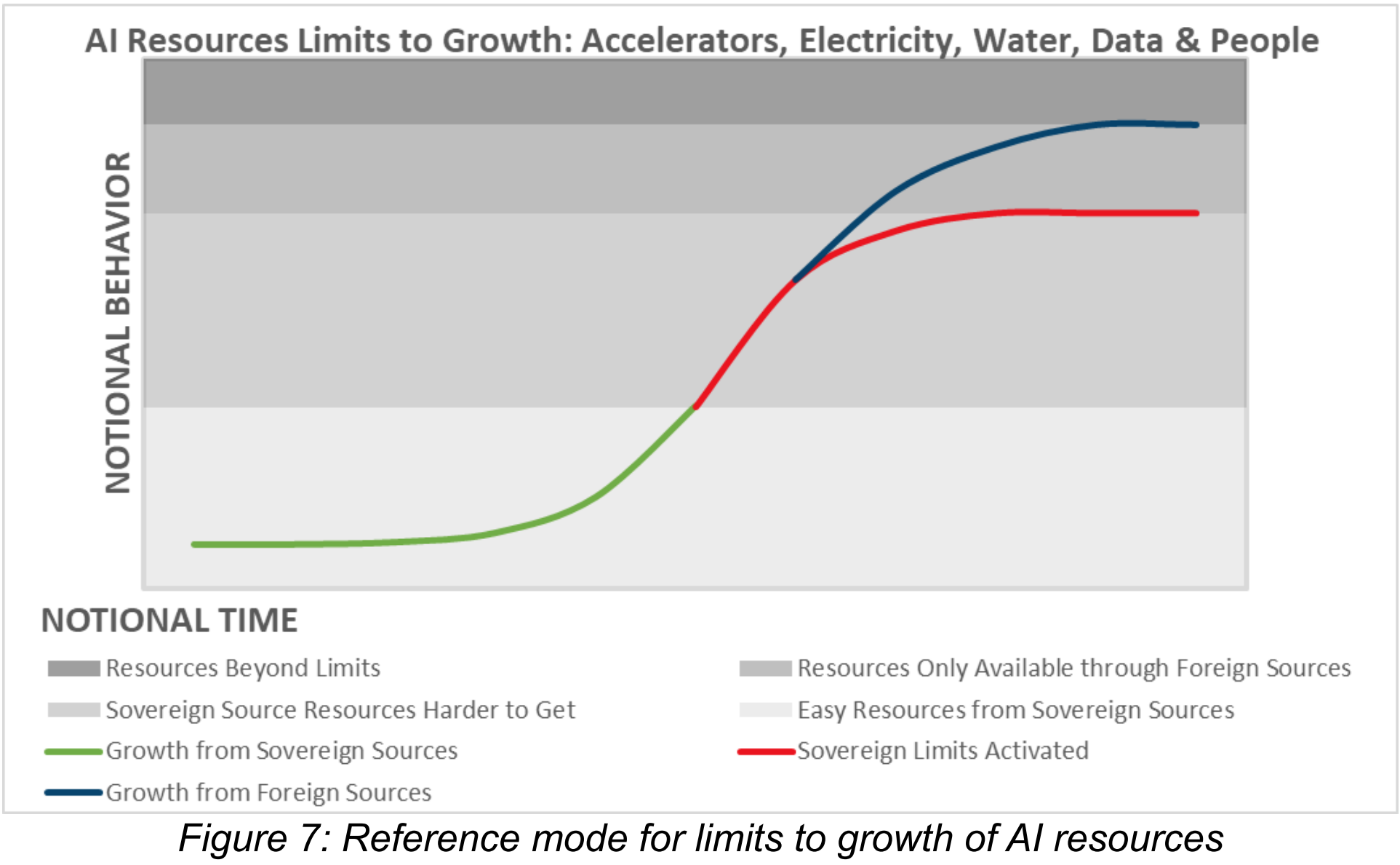

*Figure 7: Reference mode for limits to growth of AI resources*

Figure 8 shows two reference modes for the growth of AI capabilities within the focus nation, aligned with the same shaded bands in Figure 7. The left figure shows the rate of change in growth rates, while the right shows cumulative growth across use-case regions representing hoped, feared, and suspense futures. The timeline begins with rapid growth, as resources are abundant. Growth slows and stalls as those necessary resources become harder to get and then drops as resource limits are met. This generally leads to S-shaped growth in AI capabilities under the control of the focus nation.

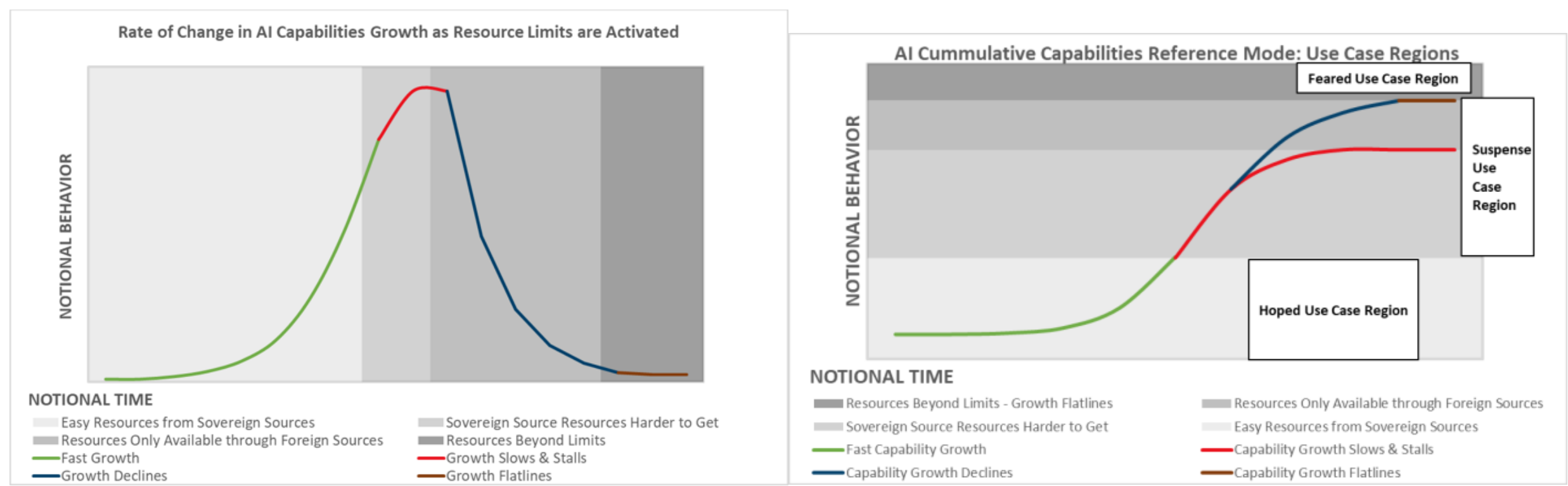

*Figure 8: Reference mode for rate of change in growth of AI capabilities (LEFT) and cumulative capabilities compared to use cases (RIGHT)*

The use-case regions help clarify where potential applications of national power might lie and how competitive dynamics might play out. Low-hanging fruit for agentic AI of automating tasks or analyzing vast amounts of data with AI agents creating their own code may be more easily accessible in the Hoped use case region of Figure 8, while over the horizon, fully autonomous sensing, understanding, and directing of national economies for optimal use may be in the Feared Use Case. Unavailable as it lies beyond the limits of resources a nation can gain for its micro, meso, and macro phenomena.

The charts in Figure 8 are also dynamic to both the focus nation represented and potential competitive dynamics. What lies in the hoped-for use-case region for a nation with plentiful access to resources may be in suspense, or what lies in the feared use-case region for a nation with significantly less resources. Likewise, strategic adversaries attempting to degrade micro- or meso-level resources, as shown in Figure 5, may cause what was in the suspense use case region to move out of reach into the feared use case.

# Discussion

Two important considerations arise from the model presented here: measures of AI sovereignty and national power and potential implications suggested by the qualitative dynamics of the model.

For over a century, nations have pursued military power as an instrument of national power in both the design and implementation of combat air power. Airpower needs brought together a technical workforce, the current body of knowledge, and experimentation in skunk works and laboratories to innovate ever more powerful capabilities and the combat airframes to contain them. Implementation focused on the ability to produce and use these aircraft as an instrument of national power, taking off and returning to airfields or aircraft carriers that housed their capability in physical, targetable locations.

Innovations in air power were labeled by generation, selecting jet aircraft during World War I as the 1st Generation and all previous innovations as precursor work. The larger the gap between generations, the less likely aircraft are to compete equitably. A nation with an air fleet of mostly 5th Generation aircraft, even if fewer in number, would fare well against a nation with more 2nd Generation aircraft. There are now six recognized generations of aircraft.

Measuring a nation's sovereign airpower required understanding its internal capability to design, produce, and field aircraft of ever-increasing generations. Even if a nation could purchase foreign aircraft, if it wasn't capable of designing and building them, it would always be reliant on a foreign partner in the supply chain. Likewise, a nation using aircraft carriers on loan or reliant on another nation's airfield could face a scenario in which access to those facilities is cut off. Based on our model, we propose notional measures for AI sovereignty and national power in agentic AI along similar lines to this understanding of combat airpower.

We selected agentic AI as the '1st Generation' of interest for these measures due to its greater capabilities and resources compared to previous iterations of GenAI, Reasoning AI, etc.

In Figure 9, we present a notional, qualitative measure of national power in agentic AI, organized into four measures. These four measures are:

1. National Power in Agentic AI Implementation (expressed in zettaFLOPS)
2. Number of Frontier Foundation agentic AI Models in the pacing measure generation
3. Number of Frontier Foundation agentic AI Models in the last three generations
4. Number of Frontier Foundation agentic AI Models across all generations

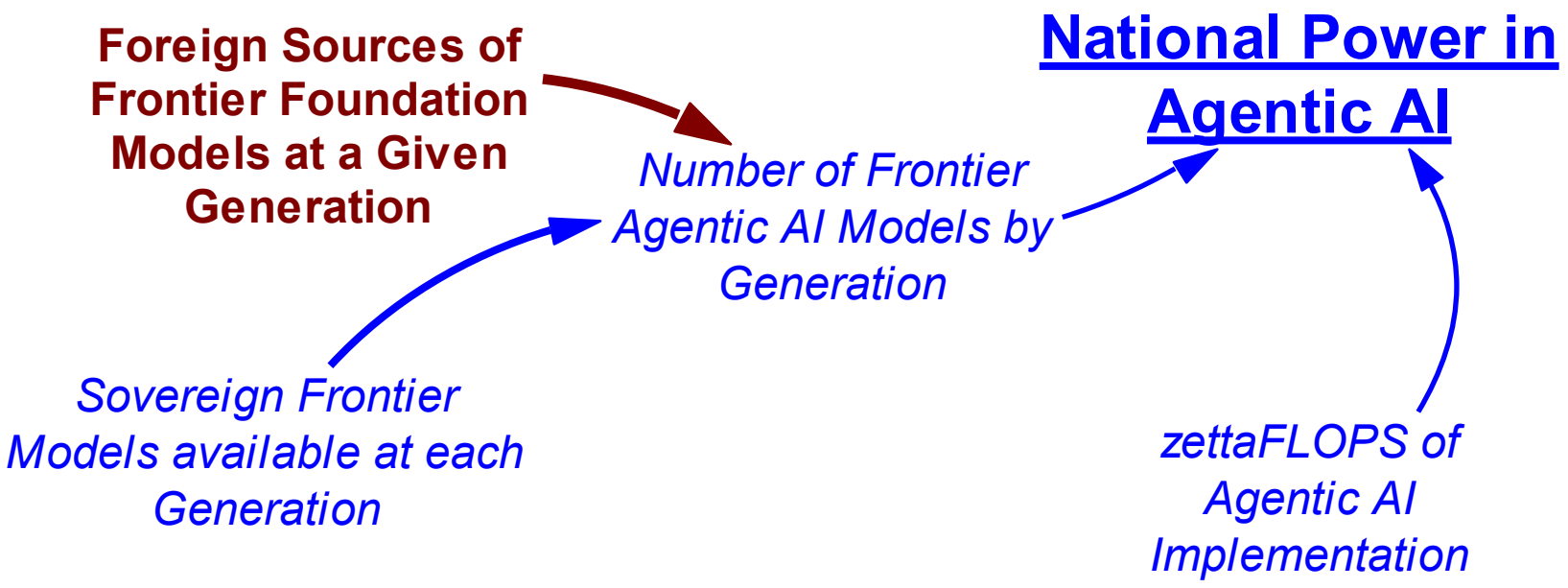

*Figure 9: National Power in Agentic AI*

These measures track the raw computed power to field agentic AI (1), the ability to maintain pace with the ever-advancing generations of new frontier models (2), a breadth of capability across the last three generations (3), and the depth of capability and experience in the total number of frontier models across all generations. A nation that can only field a few of the latest-generation combat aircraft is going to fare worse than one that can match that latest generation while also having a breadth of capabilities in recent innovations and a long, deep experience in designing, manufacturing, and fielding combat aircraft. Subsumed in this measure is both the sovereign ability to allocate resources to train new frontier models to keep pace with the innovation pacing measure or to acquire foreign frontier models that do.

For AI sovereignty, Figure 9 depicts a notional, qualitative measure as two percentages:

1. The total sovereign zettaFLOPS divided by the total zettaFLOPS.
2. The sovereign frontier models available at each generation are divided into the total number of frontier models by generation

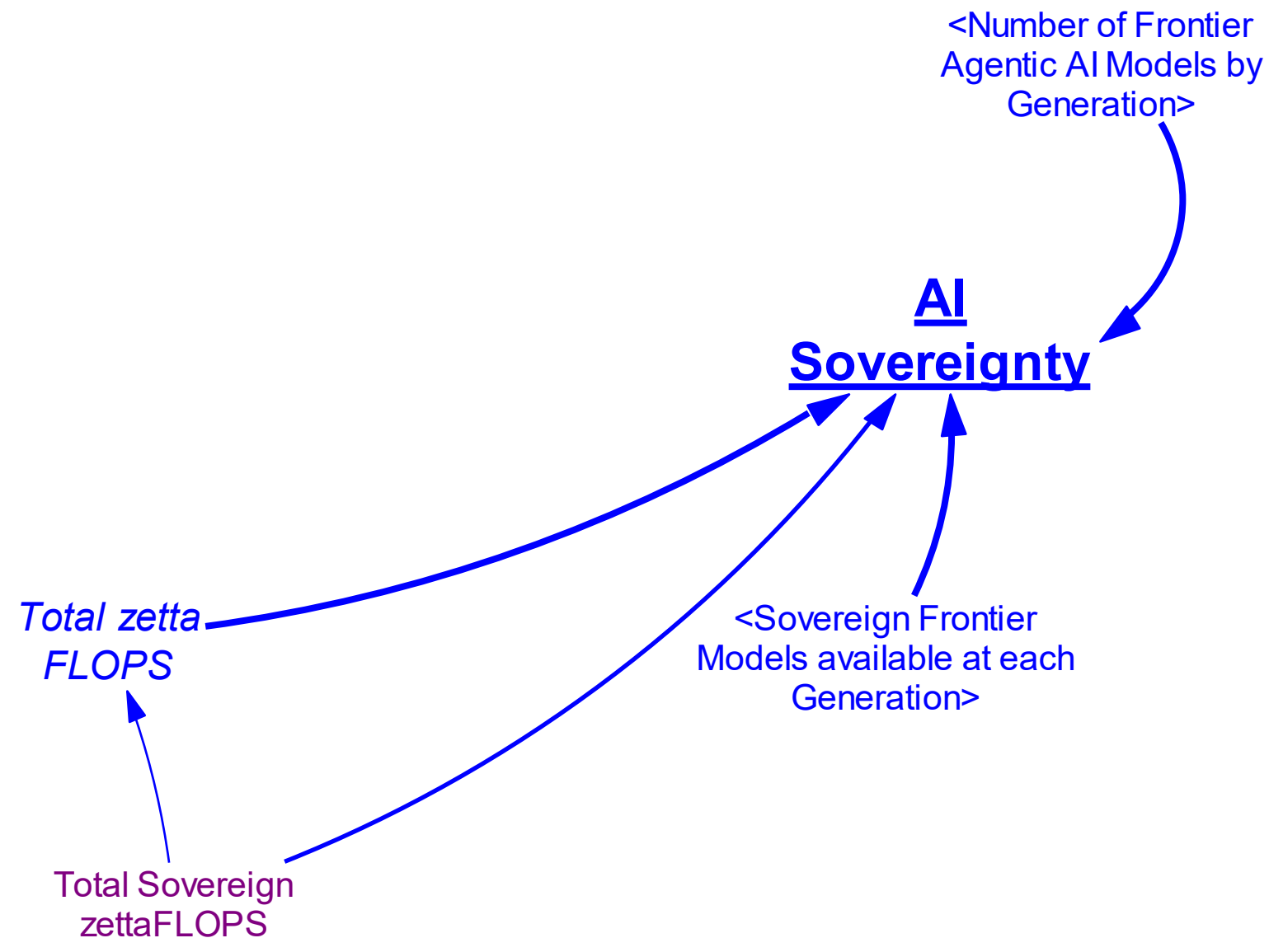

*Figure 10: AI sovereignty*

The resulting two measures, both ranging from 0 to 1, represent the percentage of overall national power that derives from sovereign sources. These measures are presented as notional and qualitative only. But in future work with computational simulation models, they can be refined and calculated.

When we began this work in January of 2026 many of these risks were notional, subject to conceptual explorations since the so-called ‘Deep Seek moment’ in late 2024[12]. But over the course of this work new risks have emerged or are accelerating and some, specifically around data centers, are worth discussing.

In the current mode of market-based AI, capabilities can seem to fade into a mesh background of distributed compute and inference. But as AI becomes an instrument of national power data centers become strategic targets, and this has already come to pass. For example, on March 1, 2026 Iran deliberately targeted two Amazon data centers in the United Arab Emirates (UAE) while debris from a downed drone in Bahrain damaged another data center[13] causing regional outages[14]. A month later Iran issued a warning against US companies it considered viable military targets in the Gulf. AI providers such as Microsoft, Google, Apple, Meta, and Nvidia were listed alongside traditional defense contractors like Boeing and GE as well as military technology firms such as Palantir[15]. Iran followed this general warning with a specific threat to target and destroy the $30B Stargate data center in UAE[16]. Although these threats did not come to fruition, the situation demonstrated the operational vulnerability of large costly data centers packed with sensitive tech to low-cost long-range drones or much more damaging ballistic missiles.

As strategic adversaries race to embed agentic AI capabilities in their militaries[17] and the mesh of distributed computing fragments into AI sovereignty, targeting key nodes and clusters of data centers training and serving military capabilities becomes plausible. In the cold war fleet trackers and target lists were maintained of aircraft frames, numbers, strategic air fields, nuclear silos and the facilities that served them. The threat vectors to these targets are not limited to kinetic strikes either. Non-kinetic effects delivered through denial measures in the domains of space, cyber or via supply chain vulnerabilities would also have to be accounted for and countered. And with community resentment to both AI in general and the rapid build out of data centers on the rise[18], information operations can seek to enhance or amplify existing discontent not just against the data centers themselves, but also against the electrical and water projects necessary to run them. Since agentic-AI can be dual use, not all targeting efforts may be aimed at military. Operations may be run to protect, or degrade, economic efforts broadly or specific niche industries including quantum, biochemical, materials, and other sectors where advances in AI are fueling research.

## Conclusion

As AI grows in capability, its implications for national security and strategic competition are likely to drive some nations to strive for AI Sovereignty. AI is likely to impact national security capabilities, economic strength, and other powers, which will increase the desirability of national AI sovereignty and increase the risk of AI dependence, analogous to historical air power dynamics. This might lead individual nations to prioritize establishing themselves as leaders in AI, which might lead to an AI arms race with not only AI technology, but the national power that it boosts as key outcomes. Strategic AI dependence relationships might also develop, with some nations declining to pursue AI sovereignty and instead utilizing AI from allied nations. These dynamics could result in major changes to the global competition and cooperation landscape.

To begin exploring this issue, this article introduced a qualitative model of AI Sovereignty and implications for national power and strategic competition. The model incorporates micro, meso, and macro resource dynamics, investment and growth dynamics, and competitive dynamics between strategic adversaries. Physical resource constraints can be significant limiting factors in the quest for AI sovereignty. The qualitative model identifies key relationships between variables, highlighting feedback loops and thus potential system behaviors. Qualitative models can also help in identifying leverage points that are likely to have significant impacts on dynamics. The CLD presented here, for example, highlights leverage points that might improve a nation's AI sovereignty or impair the AI sovereignty of strategic competitor. We also introduce measures of national power and AI sovereignty, which facilitate tracking and comparison of these key concepts.

The qualitative model introduced here is preliminary and does not represent all possible dynamics important to AI sovereignty and strategic competition. Similarly, the datasets incorporated are snapshots; we anticipate that more detailed data will be incorporated in future iterations of this work. There are also major categories of interactions that were not considered here. For example, data poisoning and AI security capabilities may have significant impacts on AI sovereignty dynamics, and the efficiency of AI may improve as technologies progress.

Future work will include developing a quantitative dynamic simulation model that integrates the qualitative model introduced here with data and mathematical representations of the key variables and their relationships. This quantitative simulation model will facilitate the investigation of resource allocation, strategic partnerships, and strategic competition. It will enable forecasting, scenario analysis, what-if analysis, and investigation of model sensitivity and the strength of leverage points. We expect this future quantitative model to illuminate key concerns about the global evolution of AI.

AI sovereignty is likely to impact not only AI development but national security, national power, and economic strength. Understanding the likely dynamics of AI sovereignty becomes more important as the influence of AI grows, and early insight into its implications will facilitate effective early actions.

# Acknowledgements

Sandia National Laboratories is a multi-mission laboratory managed and operated by National Technology & Engineering Solutions of Sandia, LLC (NTESS), a wholly owned subsidiary of Honeywell International Inc., for the U.S. Department of Energy's National Nuclear Security Administration (DOE/NNSA) under contract DE-NA0003525. This written work is authored by an employee of NTESS. The employee, not NTESS, owns the right, title and interest in and to the written work and is responsible for its contents. Any subjective views or opinions that might be expressed in the written work do not necessarily represent the views of the U.S. Government. The publisher acknowledges that the U.S. Government retains a non-exclusive, paid-up, irrevocable, world-wide license to publish or reproduce the published form of this written work or allow others to do so, for U.S. Government purposes. The DOE will provide public access to results of federally sponsored research in accordance with the DOE Public Access Plan.

## *SOURCES*

[1] A. Lee, “What Is Sovereign AI?,” Nvidia Blog. [Online]. Available: https://blogs.nvidia.com/blog/what-is-sovereign-ai/

[2] D. Acemoglu, “The simple macroeconomics of AI,” *Economic Policy*, vol. 40, no. 121, pp. 13–58, Jan. 2025, doi: 10.1093/epolic/eiae042.

[3] *Joint Warfighting*, 2023rd ed., vol. VOL 1. in Joint Publication, no. JP1, vol. VOL 1. 2023. [Online]. Available: https://keystone.ndu.edu/Portals/86/Joint%20Warfighting.pdf

[4] *DoD Dictionary of Military and Associated Terms 2021*. Department of Defense, 2021. [Online]. Available: https://irp.fas.org/doddir/dod/dictionary.pdf

[5] R. Sapkota, K. I. Roumeliotis, and M. Karkee, “AI Agents vs. Agentic AI: A Conceptual taxonomy, applications and challenges,” *Information Fusion*, vol. 126, p. 103599, Feb. 2026, doi: 10.1016/j.inffus.2025.103599.

[6] D. B. Acharya, K. Kuppan, and B. Divya, “Agentic AI: Autonomous Intelligence for Complex Goals—A Comprehensive Survey,” *IEEE Access*, vol. 13, pp. 18912–18936, 2025, doi: 10.1109/ACCESS.2025.3532853.

[7] C. Olaya, “Cows, agency, and the significance of operational thinking: Cows, Agency, and Operational Thinking,” *Syst. Dyn. Rev.*, vol. 31, no. 4, pp. 183–219, Oct. 2015, doi: 10.1002/sdr.1547.

[8] T. Clancy, S. Valeria, K. J. Oviatt, and E. Palmer, “Micro, Meso, Macro: Generic Structures of Social Complexity for Defense Systems Engineering & National Security Wargames,” in *Proceedings of the 2024 International Conference of The Computational Social Science Society of the Americas*, Cham: Springer Nature Switzerland, 2025.

[9] G. E. P. Box and N. R. Draper, *Response surfaces, mixtures, and ridge analyses*, 2nd ed. Hoboken, N.J: John Wiley, 2007.

[10] J. C. Hick and M. Izzo, “Futureproofing through 2035 for the AI and HPC Power Density Trend,” Los Alamos National Labratory, Los Alamos, Oct. 2025.

[11] A. Souly *et al.*, “Poisoning Attacks on LLMs Require a Near-constant Number of Poison Samples,” 2025, *arXiv*. doi: 10.48550/ARXIV.2510.07192.

[12] T. Clancy, “MegaMullet: The DeepSeek Moment – The Start of an AI Cold War,” InfoMullet. [Online]. Available: https://infomullet.com/2025/02/12/deepseek_moment/

[13] D. Boffey, “‘It means missile defence on datacentres’: drone strikes raise doubts over Gulf as AI superpower,” *The Guardian*, Mar. 07, 2026. [Online]. Available: https://www.theguardian.com/world/2026/mar/07/it-means-missile-defence-on-data-centres-drone-strikes-raises-doubts-over-gulf-as-ai-superpower

[14] S. Moss, “Amazon confirms two UAE data centers hit by drone strikes, third in Bahrain damaged,” *Data Center Dynamics*, Mar. 03, 2026. [Online]. Available: https://www.datacenterdynamics.com/en/news/amazon-confirms-two-uae-data-centers-hit-by-drone-strikes-third-in-bahrain-damaged/

[15] D. Alomar and C. Sertin, “Iran Warns US Tech Firms And UAE AI Company G42 Will Be Targeted on 1 April.” [Online]. Available: https://www.wired.me/story/war-on-big-tech-iran-names-israeli-linked-us-firms-as-potential-targets

[16] S. Moss, “Iran threatens to attack OpenAI’s Stargate data center in UAE,” *Data Center Dynamics*, Apr. 07, 2026. [Online]. Available:

https://www.datacenterdynamics.com/en/news/iran-threatens-to-attack-openais-stargate-data-center-in-uae-war/

[17] S. Frenkel, P. Mozor, and A. Satarino, “Mutually Automated Destruction: The Escalating Global A.I. Arms Race,” *New York Times*, New York Times, Apr. 12, 2026. [Online]. Available: https://www.nytimes.com/2026/04/12/technology/china-russia-us-ai-weapons.html

[18] A. Ramkumar, K. Blunt, and L. Ellis, “The American Rebellion Against AI Is Gaining Steam,” *Wall Street Journal*, May 19, 2026. [Online]. Available: https://www.wsj.com/tech/ai/the-american-rebellion-against-ai-is-gaining-steam-94b72529

## *SOURCES IN SUPPLEMENTARY*

[1] G. Hamm, “2024 United States Data Center Energy Usage Report,” Lawrence Berkely National Laboratory, 2025. doi: 10.71468/P1WC7Q.

[2] “Frontier Data Centers,” Epoch AI. [Online]. Available: https://epoch.ai/data/data-centers

[3] D. Owen *et al.*, “Notable AI Models Dataset.” Feb. 12, 2026. Accessed: Feb. 14, 2026. [Online]. Available: https://epoch.ai/data/ai-models-documentation#downloads

[4] A. Laurent, “NVIDIA HGX Platform: Data Center Physical Requirements Guide,” Jan. 2026. [Online]. Available: https://intuitionlabs.ai/articles/nvidia-hgx-data-center-requirements

[5] S. Cruzes, “DATA CENTERS IN THE AGE OF AI: A TUTORIAL SURVEY ON INFRASTRUCTURE, SUSTAINABILITY, AND EMERGING CHALLENGES,” Oct. 27, 2025, *Preprints*. doi: 10.36227/techrxiv.176158592.23065552/v1.

[6] “Glossary of Data Center Terms.” [Online]. Available: https://www.pducables.com/resources/glossary-of-data-center-terms

[7] “NVIDIA DGX SuperPOD: Data Center Design Featuring NVIDIA DGX H100 Systems.” [Online]. Available: https://docs.nvidia.com/dgx-superpod/design-guides/dgx-superpod-data-center-design-h100/latest/index.html#dgx-superpod-data-center-design-featuring-dgx-h100

[8] “How many servers does a data center have?,” RackSolutions. [Online]. Available: https://www.racksolutions.com/news/blog/how-many-servers-does-a-data-center-have

[9] J. C. Hick and M. Izzo, “Futureproofing through 2035 for the AI and HPC Power Density Trend,” Los Alamos National Labratory, Los Alamos, Oct. 2025.

[10] A. Katal, S. Dahiya, and T. Choudhury, “Energy efficiency in cloud computing data centers: a survey on software technologies,” *Cluster Comput*, vol. 26, no. 3, pp. 1845–1875, Jun. 2023, doi: 10.1007/s10586-022-03713-0.

[11] D. Mytton, “Data centre water consumption,” *npj Clean Water*, vol. 4, no. 1, p. 11, Feb. 2021, doi: 10.1038/s41545-021-00101-w.

[12] E. Keski-Nisula, “Demand response potential of AI data center facilities — Physical flexibility of cooling, energy storage, and backup power in Finnish wholesale and reserve electricity markets,” Aalto University School of Electrical Engineering, 2025.

[13] A. Walters, A. Walker, and M. Kelley, “Rethinking AI Demand Part 1: AI Data Centers Are Experiencing a Surge of Training Demand - What Happens When the

Surge is Over?” [Online]. Available: https://www.alvarezandmarsal.com/insights/rethinking-ai-demand-part-1-ai-data-centers-are-experiencing-surge-training-demand-what

[14] B. Srivathsan, M. Sorel, and P. Sachdeva, “AI power: Expanding data center capacity to meet growing demand,” McKinsey. [Online]. Available: https://www.mckinsey.com/industries/technology-media-and-telecommunications/our-insights/ai-power-expanding-data-center-capacity-to-meet-growing-demand#/

[15] B. Cottier, R. Rahman, L. Fattorini, N. Maslej, T. Besiroglu, and D. Owen, “The rising costs of training frontier AI models,” 2024, *arXiv*. doi: 10.48550/ARXIV.2405.21015.

[16] J. Sevilla *et al.*, “Can AI scaling continue through 2030?,” Epoch AI. [Online]. Available: https://epoch.ai/blog/can-ai-scaling-continue-through-2030

[17] A. Souly *et al.*, “Poisoning Attacks on LLMs Require a Near-constant Number of Poison Samples,” 2025, *arXiv*. doi: 10.48550/ARXIV.2510.07192.

[18] “Our World in Data - Artificial Intelligence,” Our World in Data.

# AI Sovereignty and National Power
# Supplementary Materials

## Contents

The supplementary materials below provide additional notes on the structure of the model including micro, meso, and macro parameters and how they could be used in the mode.

Recent data values are provided with most parameters. A caveat is provided however that the current value of any parameter in a fast-changing technology like agentic AI are constantly shifting and will probably be out of date at time of publishing. However, the structural relationship between micro, meso, and macro parameters and their location in the model are likely to be more enduring. In a computational simulation the values of any given parameter can be set to update regularly providing for the most current value.

# Micro Parameters: Accelerators, Servers, & AI Server Cabinets

## *Measures of Compute: FLOP, petaFLOPS, exaFLOPS, or zettaFLOPS and /s-days*

Compute power is expressed both as a quantity and a performance, or in system dynamics terms as a stock and a flow. A value and a rate of change. The standard quantity measure of compute power is a 'floating point operations' (FLOP). A quadrillion (1.0E+15) FLOP are known as a petaFLOP which historically was a more useful unit of measure for training and data needs. The standard unit of performance, or rate of computations, was then a petaFLOP per second, expressed in this paper as petaFLOPS. This is the ability of an accelerator to conduct one petaFLOP of compute for one second. As there are 86,400 seconds in a day a 1 petaFLOPS-day is equivalent to 86,400 petaFLOP. As compute rates have rapidly grown however, larger denominations are more useful, especially for modeling national levels of compute. For quantity measures there are 1,000 petaFLOP is an exaFLOP and a 1,000 exaFLOP is a zettaFLOP. These larger units have the same notations for performance measures (e.g. exaFLOPS, zettaFLOPS/day).

## *Generators of Compute: Accelerators*

To generate these levels of compute are specialized accelerators which "allow the server to more quickly process large quantities of calculations in parallel" that may be based on GPU, ASIC, or TPU chips[1, p. 16]. We select the Nvidia DGX H100 8-GPU accelerator as a baseline for this work given its prevalence in literature and our datasets [1, p. 22], [2], [3]. As hardware changes rapidly, even as we write this paper H100 is approaching the end of life. Yet it still challenges legacy data center infrastructure in ways that illuminate achieving AI Sovereignty and National Power in Agentic AI may not be realized simply by repurposing decades old data center technology[4].

## *Compute for AI Training*

In our conceptual analogy Training AI is likened to the design and prototyping of new aircraft frames and capabilities. At a high level AI training involves long-duration is "intensive, long-duration computation with relatively predictable power draws, typically performed in dedicated facilities with specialized hardware [1, p. 22]."

We measure Training AI as the required compute to train a model (FLOP), the size of the data set being trained on (FLOP), and the time it takes to train in hours or days to train. This ignores additional fine tuning, alignment, and other considerations but provides a useful rough order, especially as the difficulty of training each iterative generation of model increase exponentially meaning the long tail of other adjustments are captured within exponential growth of the three measures we identify.

## *Compute for AI Implementation*

Once trained, an AI model can serve inferences or be 'implemented' at scale, be likened in our conceptual analogy to the use of the aircraft frames in daily operations to affect an instrument of national power. As serving inference is based on prompts by users and varies in size based on the complexity of tasks, implementation "presents more variable loads across a heterogeneous hardware mix, making it more challenging to model but increasingly important as deployment scales[1, p. 22]." Although implementation used to be cheap in terms of compute for GenAI, the evolution to Agentic AI with its longer context lengths and far more complex operations is approaching the dense high intensity requirements of Compute for AI Training[5, p. 23].

## *Physical Units of Compute: Tiles, Cabinets & Data Center Space*

### Tiles

In a data center the physical space available for computing power is allocated to 'tiles'. Each tile is 2'x2' of raised floor mounted above the actual floor to allow cabling and cooling to access the computer hardware above[6].

### Cabinet

On these tiles are located 'cabinets which is the physical housing for servers. Convention dictates the measurement of a server cabinet is on the interior of the housing and relate to the size of available for placing standardized rack units. U stands for a space capable of housing a shelf of 1.75” of vertical space, so a 42U server rack contains 73.5” of usable computing space on the inside and is nearly 7’ on the outside.

**Cabinet Capacity for AI Servers**

A standard cabinet described above, as a best practice, can hold four AI Servers each mounted on individual racks configured with Nvidia DGX H100, each consisting of 8 GPUs providing a total of 32 GPU’s[4]. As each Nvidia GGX H100 can provide 4 petaFLOPS of compute, each rack of 8 GPUs can generate 32 petaFLOPS and the cabinet can generate 128 petaFLOPS[4]. The rest of the space inside the cabinet, and the tile footprint is taken up by extensive cooling, network controls, and even white space around the cabinets to allow proper heat dispersion[4], [7].

**Tile Usage per AI Cabinet**

For all needs inside and outside a computer cabinet Nvidia recommends 32” x 48” of space[7], or the use of a 2x2 tile configuration.

**Determining Data Center SqFt from Tiling & Occupied**

Although exact percentages vary, heuristic rules of thumb are that 64% of available tiles are occupied by server cabinets with the remaining going vacant for future expansion, space, heat or suboptimal distance for clustering. All the tiles, occupied and unoccupied, are known as the “white space” of a data center. White space is on average 65% of a data center’s total space, with the remaining taken up by “gray space” of networking, cabling, cooling and other infrastructure outside the rack components [8].

**AI Cabinet Annual Electricity & Growth Unit: kW**

AI Server electricity needs vary by their use in training or implementation and how those uses may require mode-shifting between rated, maximum, operational and idle [1, p. 17]. Although this is commonly called ‘power draw’, we use ‘electricity draw’ instead so as not to confuse the reader with our use of ‘power’ in the context of National Power and compute power. Annual electricity needs “The average power draw over an entire year, considering all operating modes and power levels[1, p. 17].” Current power per rack needs for AI Servers are between 30kw-100kw[5, p. 2] up to 250kw[9, p. 4] per rack. The logarithmic plot in Figure 1 shows a behavior mode in the average data center rack load and peak performance rack power requirements historically from 2000-2024 and future forecasts through 2034[9, p. 5]. For AI Training Data Centers the power density is more likely to be 100 kW or higher per rack while AI Inference Data Centers use power densities on the lower range of 15-30 kW [5, p. 4]. With four AI server racks per cabinets, these estimates are all multiplied x4 to achieve the total AI Cabinet need in the model. In both the current and future forecast, the power draw of AI servers used in inference delivery and training are held constant at 40% and 80% operational up time respectively[1, p. 27],

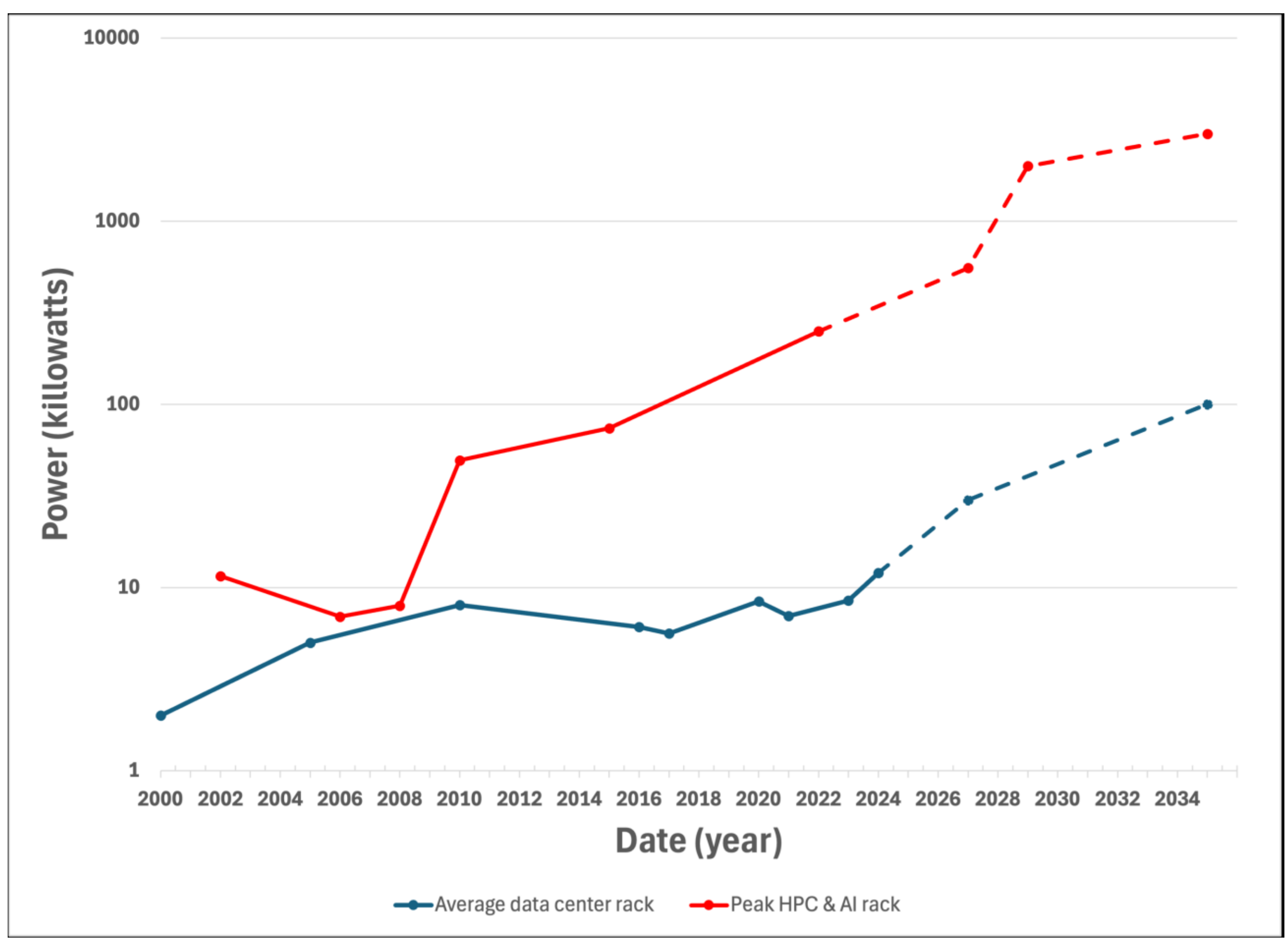

*Figure 1: Lawrence Berkely National Laboratory Historical and Forecasted Averaged and Peak Power Loads per Rack*

## **AI Cabinet Water Needs**

Water needs of AI Cabinets is directly related to the power density per rack. Although there are numerous non-water cooling strategies that can be employed at rack power densities below 100kw above that mark liquid cooling is a requirement[9, p. 2], [5, p. 10]. Direct to Chip (D2C) cooling can support 75-175 kw per rack while liquid immersion cooling can support 150-225 kW racks[5, p. 14].

## **AI Cabinet Accelerators**

Using the Nvidia guidance of no more than four DGX H100 servers deployed in a cabinet there would be 32 accelerators per cabinet. [4]

.

# Meso Parameters: Data Centers

## ***Power Usage Efficiency (PUE) Unit: kWh/kWh***

Measures data-center electricity efficiency by taking "total facility power split by IT hardware[10, p. 1846]." While a PUE of 1.5 means that 1.5MW would have to reach the data center for 1MW to be used by the computer components[5, p. 9]. Though PUE is not the only measure, it is a simplified way of measuring all loses of power to other sources ranging from administrative functions to transformer losses, which can account for up to 10% of all usage [9, p. 7]. The average annual PUE for data centers was 1.57 PUE in 2021[10]. The specific average PUE on AI Specialized data centers is 1.14 [1, p. 47]

## ***Emissions Unit: kw***

Emissions are calculated based on the kWh of electricity at the source with a factor of 0.35 kg/kWh of $CO_2$ equivalent[1, p. 57].

## ***Water Usage Efficiency (WUE) Unit: liters/kWh***

Similar to PUE, water usage efficiency (WUE) measures water usage at data centers. Direct or onsite WUE measures the water usage at the data center itself, used in cooling, while source WUE includes water usage in the generation of power [1, p. 39]. The average annual WUE is the annual site water usage used at the center divided by the total IT equipment energy with units of Liters per KWh (L/kWh)[11, p. 11]. Note this measure only accounts for water used in the cooling of servers and does not include water that may be required for upstream generation of power, which is outside the boundary of our model. The average direct WUE on the four types of data centers we model are Small (.32), Midsize & Colo (.67), Hyperscale (.32), and AI Specialized (.61) [1, p. 47].

## ***# of Data Centers by Type:***

There is no standard terminology or taxonomy of data centers and different labels include small, midsize, colo, internet, and hyperscale with some existing hyperscale data centers drawing 5-100 MW of electricity [12, p. 31]. However, because of our choice to model the 1$^{st}$ Generation of National Power at the level of Agentic AI, which requires far more compute to train and implement than previous versions, we can focus on data centers designed for concentrating AI accelerators with corresponding electricity density and liquid cooling requirement. For our purposes, we track AI Specialized Data Centers which may have the MW capacity of a hyperscale, but is distinguished by its concentration of AI accelerators and corresponding high electricity density per rack requiring water cooling in addition to electricity[12, p. 36], a feature which requires design in the architecture of the data center itself and is difficult to retrofit.

We used a dataset of open source-derived characteristics of modern US and Chinese AI 'frontier' data centers to provide rough estimates of AI specialized data centers. The Epoch AI report lists 19 projects, which when entries with blank values for H100 equivalent compute, power, and capital cost are removed leaves 11 AI specialized data center listed below. In Table 1 we report the Epoch AI findings of current H100 accelerator equivalents, the direct power needs and known total capital costs. Project handle is a combination of sponsor, location, and project name.

*Table 1: Epoch AI OpenSource Findings on Frontier AI Data Center Projects*

| Handle | Current H100 equivalents | Current direct power (MW) | Current total capital cost (2025 USD billions) |
|---|---|---|---|
| Google Omaha Nebraska | 109,880 | 189 | 6.05 |
| Google Pryor Oklahoma | 62,851 | 195 | 6.21 |
| xAI Colossus 2 Memphis Tennessee | 280,838 | 279 | 8.90 |
| OpenAI-Oracle Stargate Abilene Texas | 254,674 | 295 | 8.00 |
| Amazon Canton Mississippi | 214,348 | 341 | 10.89 |
| Alibaba Zhangbei Zhangjiakou Hebei | 168,128 | 474 | 15.12 |
| xAI Colossus 1 Memphis Tennessee | 275,796 | 498 | 12.88 |
| Microsoft Fairwater Fayetteville Georgia | 436,569 | 506 | 13.84 |
| Google New Albany Ohio | 235,426 | 543 | 17.34 |
| Meta Prometheus New Albany Ohio | 475,704 | 614 | 19.60 |
| Anthropic-Amazon Project Rainier New Carlisle Indiana | 471,566 | 751 | 23.97 |

Using this open source data and the researched parameters described above the authors derived additional data center values useful for our simulation. In Table 2 we use the H100 equivalents provided by Epoch AI to derive total zettaFLOPS of compute, the source power required using PUE and the direct water cooling needs of these centers.

*Table 2: Authors Estimated Values on Frontier AI Data Center Projects for Compute, Power, and Water*

| Project Handle | Current H100 equivalents | Est. Total Data Center Compute Generated at 4 petaFLOPS per H100 | Est. Total Data Center Compute Generated in zettaFLOPS | Est. Source Power with PUE of 1.14 (MW) | Est. Water with WUE of .61 (Liters) |
|---|---|---|---|---|---|
| Google Omaha Nebraska | 109,880 | 439,520 | 0.44 | 215.46 | 115,290 |
| Google Pryor Oklahoma | 62,851 | 251,404 | 0.25 | 222.3 | 118,950 |
| xAI Colossus 2 Memphis Tennessee | 280,838 | 1,123,352 | 1.12 | 318.06 | 170,190 |
| OpenAI-Oracle Stargate Abilene Texas | 254,674 | 1,018,696 | 1.02 | 336.3 | 179,950 |
| Amazon Canton Mississippi | 214,348 | 857,392 | 0.86 | 388.74 | 208,010 |
| Alibaba Zhangbei Zhangjiakou Hebei | 168,128 | 672,512 | 0.67 | 540.36 | 289,140 |

| xAI Colossus 1 Memphis Tennessee | 275,796 | 1,103,184 | 1.10 | 567.72 | 303,780 |
|---|---|---|---|---|---|
| Microsoft Fairwater Fayetteville Georgia | 436,569 | 1,746,276 | 1.75 | 576.84 | 308,660 |
| Google New Albany Ohio | 235,426 | 941,704 | 0.94 | 619.02 | 331,230 |
| Meta Prometheus New Albany Ohio | 475,704 | 1,902,816 | 1.90 | 699.96 | 374,540 |
| Anthropic-Amazon Project Rainier New Carlisle Indiana | 471,566 | 1,886,264 | 1.89 | 856.14 | 458,110 |

In Table 3 we use the H100 equivalents provided by Epoch AI to derive total AI Server Cabinets, required tiles and from that calculate the overall SqFt and km^2 footprint of the data centers.

*Table 3: Authors Estimated Values on Frontier AI Data Center Projects for Space*

| Project Handle | Current H100 equivalents | Est. Cabinets | Est. Tiles | Est Total SqFt of Data Center | Est Total km^2 of Data Center space |
|---|---|---|---|---|---|
| Google Omaha Nebraska | 109,880 | 3,434 | 13,735 | 136,260 | 0.01 |
| Google Pryor Oklahoma | 62,851 | 1,964 | 7,856 | 77,940 | 0.01 |
| xAI Colossus 2 Memphis Tennessee | 280,838 | 8,776 | 35,105 | 348,261 | 0.03 |
| OpenAI-Oracle Stargate Abilene Texas | 254,674 | 7,959 | 31,834 | 315,816 | 0.03 |
| Amazon Canton Mississippi | 214,348 | 6,698 | 26,794 | 265,809 | 0.02 |
| Alibaba Zhangbei Zhangjiakou Hebei | 168,128 | 5,254 | 21,016 | 208,492 | 0.02 |
| xAI Colossus 1 Memphis Tennessee | 275,796 | 8,619 | 34,475 | 342,009 | 0.03 |
| Microsoft Fairwater Fayetteville Georgia | 436,569 | 13,643 | 54,571 | 541,380 | 0.05 |
| Google New Albany Ohio | 235,426 | 7,357 | 29,428 | 291,947 | 0.03 |
| Meta Prometheus New Albany Ohio | 475,704 | 14,866 | 59,463 | 589,911 | 0.05 |
| Anthropic-Amazon Project Rainier New Carlisle Indiana | 471,566 | 14,736 | 58,946 | 584,779 | 0.05 |

To gain SqFt of Data Space we estimated the number of AI Server Cabinets necessary to serve the H100 equivalents. Given the standard tile requirements and standard tile size from AI Server Cabinets, we derived the white space these server cabinets would occupy. White space tiling occupied with servers usually only accounts of 64% of the data center white space, allowing us to calculate a total white space footprint, and then using a known ratio calculate the remaining gray space of the center resulting in the total data center SQFT.

Due to the enormous weight of servers and cooling, most data centers have traditionally been a single story. However, some data centers are experimenting with two, three and even more floors in height to optimize space and reduce latency across the clusters.

The data center space estimates then may not represent the total footprint on the ground, but rather the interior space required.

## Macro Parameters: National Power in Agentic AI and AI Sovereignty

Marco parameters influence the level of AI sovereignty and national power described in the main article contrasted with meso (data center) or micro (AI Server Cabinet or even Racks) parameters.  because training and implementation have different operational compute factors, and the allocation between them in our model is determined at the national level.

### *Operational & Delivered Compute*

The operational compute power takes the operational uptime power draw from above sources, 40% for inference implementation and 80% for training, and multiplies that by the allocation for inference or training at the national level. So, 1 petaFLOPs-day of inference compute capability would result in .4 petaFLOPS-day in operational compute.

Delivered compute is operational compute further constrained by the so-called ‘compute-memory gap’. In addition to compute power generated by accelerators, AI Servers require memory to store the information being processed. And the growth of compute power in FLOP is exceeding the growth rate in the speed of memory, resulting in the so-called ‘compute-memory gap.’ Though beyond the scope of this paper the difference between theoretical FLOP compute and delivered FLOP compute may range between 1%-30%. For this paper we assume a parameter of 10%. In the operational compute example 1 petaFLOPs-day of inference compute capability would result in .4 petaFLOPS-day in operational compute that results in 0.4 petaFLOPS-day in delivered compute.

To validate that our calculations on operational and delivered compute are notionally useful for the problem we are addressing, we reverse engineered the training hours of notable AI models compared to reported results. To do this, we filtered Epoch AI’s dataset of Notable AI models to those published in 2024 or later, that had listed training FLOP, training time in hours, accelerator type, and quantity of accelerators [3]. When comparing our calculated training hours to the reported and scaling it our estimates ranged from .69 to 1.76 of reported with a mean of 1.05 and a median of .96.  (See Supplementary for calculation sheet.)

### *Allocation to Training: Training vs. Inference Implementation Split*

Compute power in AI is split between training new models and inference or serving implementation of already trained models. Both are important measures. Although new capabilities are unlocked through training, the potential of those capabilities is only delivered through inference serving and implementation. Training new models recently required nearly 70% of AI Data Center capacity; however this is expected to shift over the next 5-10 years as frontier foundation Agentic AI models are trained, and the workload to shift to AI Inference[13],[14]. The allocation to training parameter in the proposed model is a fraction that can be multiplied against Total zettaFLOPS to determine the compute to training, or as (1-Allocation for Training) to determine the national zettaFLOPS allocated for inference and implementation.

### *Pacing Measure: Training Frontier Agentic AI Foundation Models and at successive Generations of Capability*

Given our purpose to forecast both National Power and Sovereignty in Agentic AI, being able to forecast a pacing measure is key. The pacing measure for our work is the anticipated future needs to develop successively more powerful frontier AI models. Broadly, we conceptualize the pacing measure as consisting of three components related to the scaling hypothesis (1-3 below) and an additional measure of the workers needed to oversee the process (4 below):

1. Required training compute
2. Required training hours
3. High quality data set size
4. Number of highly skilled technical workers

Although this simplifies a vast amount of complexity, for a pacing measure at the level of national power, we believe it is useful to start and can be further improved upon.

New frontier foundation models are then accumulated to unlock successive generation of agentic AI, though at this stage of research that number is arbitrarily chosen.

In the following sections we first examine the historical behaviors of the scaling hypothesis components of the pacing measure as well as the technical workforce growth. From this we establish a highly abstracted aggregate baseline pacing measure for 2026 and future growth factors. We note these findings are extremely preliminary and should be taken as notional and directional at the level of national power in agentic AI, rather than specific for any given frontier foundation model.

For most analysis and charts below, except where noted, we rely on Epoch AI’s Notable Model dataset [3]. Epoch AI also has a Frontier AI Model dataset tracking the top ten models by training compute at time of release. However, at best the sample size of each year’s models is only ten, and many fields of interest are marked with ‘unspecified’ or ‘unreleased.’ Epoch AI’s recommends using it’s Notable AI dataset for analysis. Each year hundreds of models are included in that dataset, selected by Epoch AI meeting notability criteria. This introduces a risk that our estimated parameters may be lower than actual values, but these can be adjusted upward easily as more data becomes available.

## Historical Behavior of Scaling Hypothesis Pacing Measure Components: Training Compute, Training Time, and Data Set Size

Our analysis of the historical growth in pacing measures of Training Compute (FLOP), Training Time (Hours), and Training Data Set Size (FLOP) in Figure 2, Figure 3, and Figure 4. We arbitrarily selected 2010 as the starting point as the first year that had values in all three parameters. For each category we calculated max, min, average and count of models in the sample size, since not all records in the Notable AI Dataset always have the same data [3].

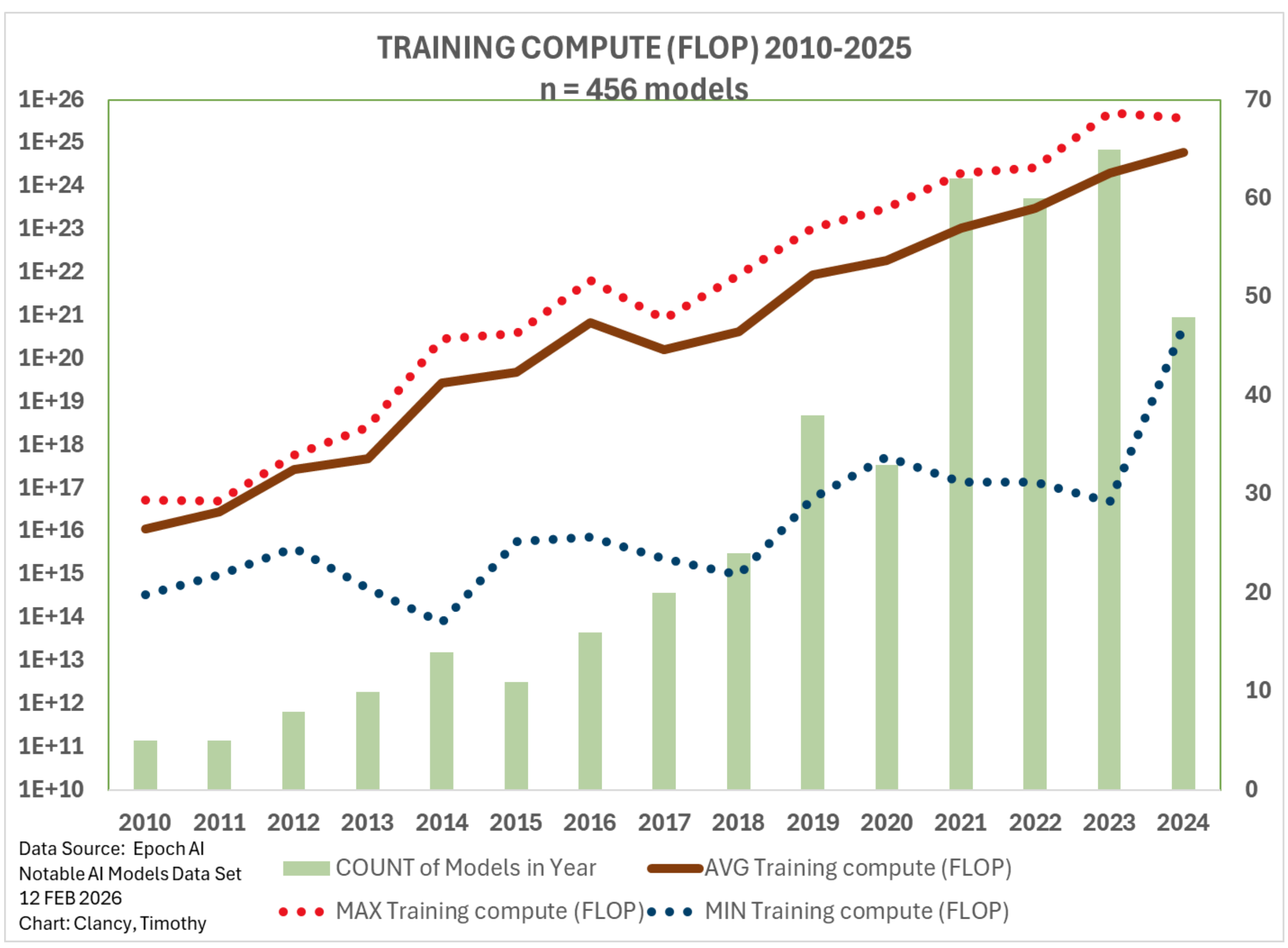

*Figure 2: Authors analysis of Epoch AI OpenSource Notable Models Findings on Training Compute*

Using hardware acquisition increases to approximate training growth requirements results in a range of 2.4-3.0x growth in training requirements since 2016 and a historical doubling time of 8-9 months[15, p. 5]. Data centers in the 1-5 GW range can support training runs from 1e28 to 3e29 FLOP while distributed data centers with aggregate 2 – 45 GW can support 2e28 to 2e30 FLOP, with a high end of 3e29 to 2e31 FLOP[16].

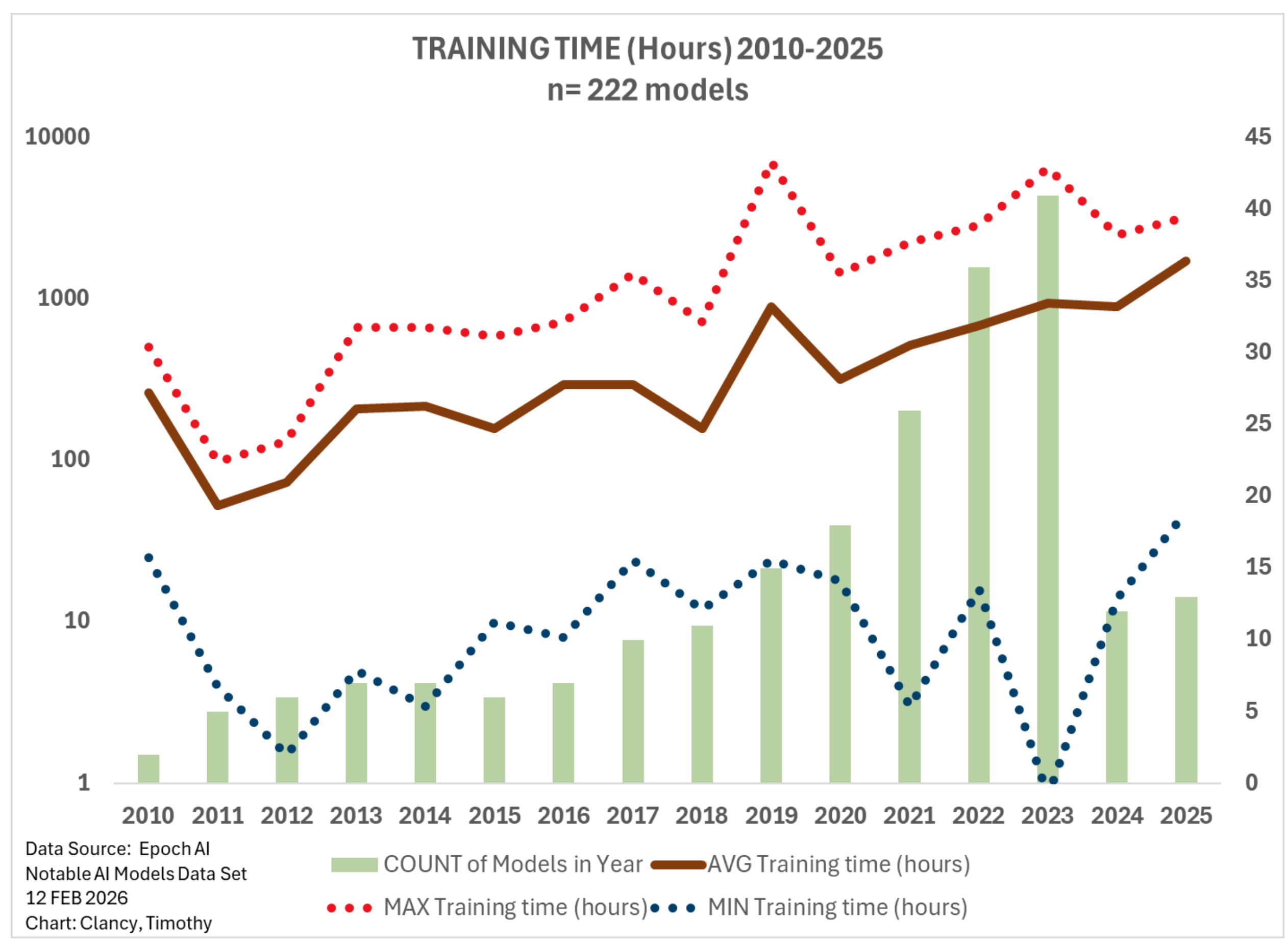

*Figure 3: Authors analysis of Epoch AI OpenSource Notable Models Findings on Training Time*

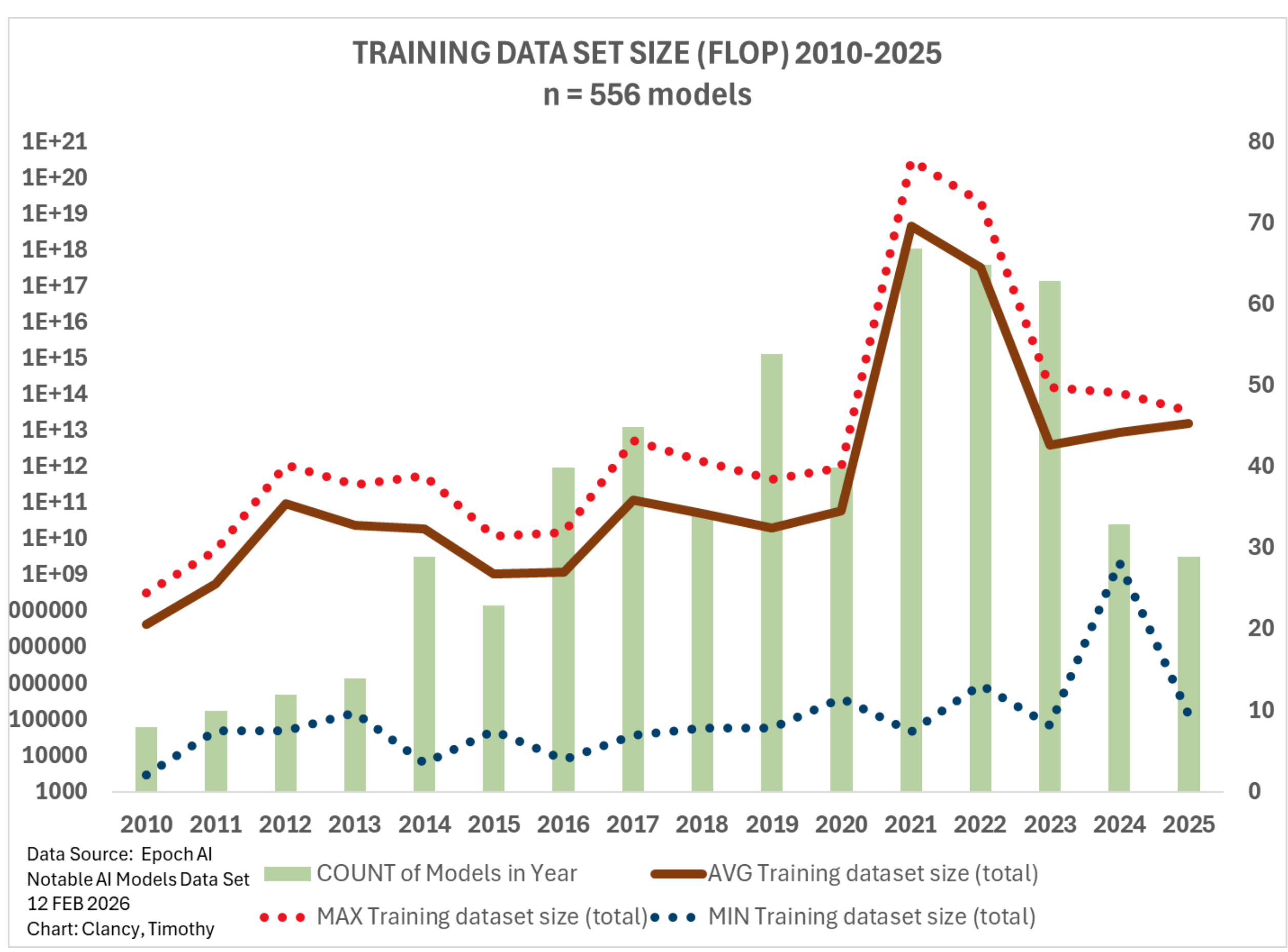

*Figure 4: Authors analysis of Epoch AI OpenSource Notable Models Findings on Training Data Set Size*

Data sets are obtained from the total tokens available to train, converted into the FLOP of compute power to train. For example, Epoch AI estimates that total token availability for data training across text, multi-modal video, image, and audio will be 400 trillion to 20 quadrillion tokens, or 6e28 FLOP to 2e32 FLOP training[16]. This total data set availability is not the same as high-quality data sets. The actual amount of training data available may be reduced based on the frontiers model purpose and filtering out corrupted, or poisoned data sets[17].

## Historical Behavior of Technical Workforce Growth

The technical workforce for Agentic AI are personnel who support both major operations: Training of models and Implementation through the serving of inference. The former workforce is much smaller than the latter. Designing and training Frontier models is a skillset possessed by so few that those highly skilled at it have been highly sought after gaining compensation packages of hundreds of millions of dollars to billions of dollars. But they are the tip of a very large pyramid of skilled workers necessary to serve the models, maintain the data centers, and integrate them into as an instrument of national power across industrial, business, and national security uses. Measuring this technical workforce is challenging in AI. Many programmers forgo completing formal education at the undergraduate, graduate, or PhD level instead opting to directly enter the workforce. Proxy measures are only roughly useful, such as the percentage of AI jobs among all new job openings, scholarly publications by country, or the attendance at AI conferences as reported in Figure 5, Figure 6, and Figure 7 respectively obtained from Our World in Data[18].

**Share of artificial intelligence jobs among all job postings**

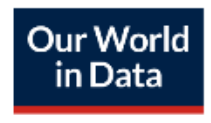

A job posting is considered an AI job if it requests one or more AI skills, e.g., "natural language processing", "neural networks", "machine learning", or "robotics".

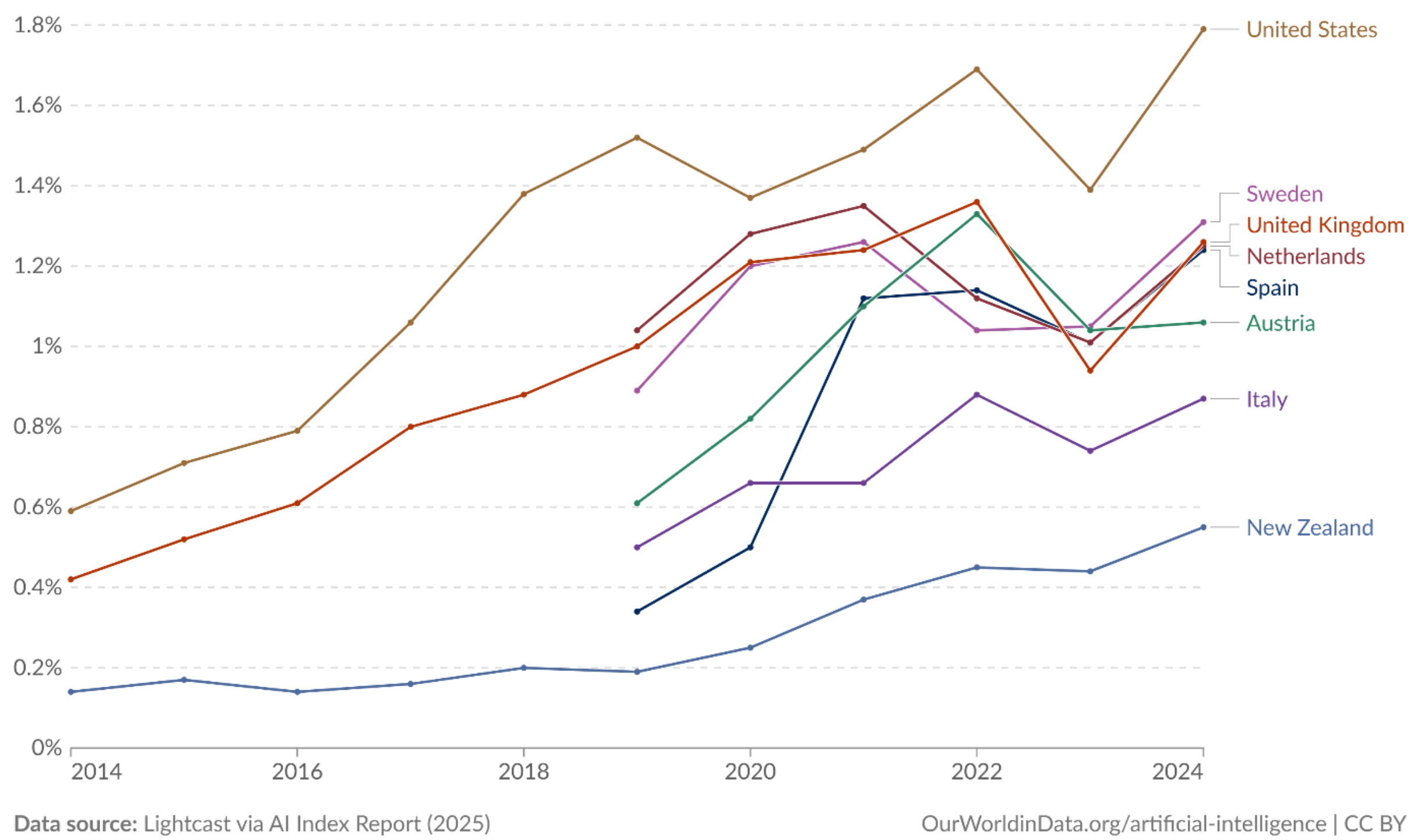

*Figure 5: Percentage of AI Jobs among all Job Postings from Our World in Data*

## Annual scholarly publications on artificial intelligence, 2023

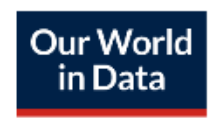

English- and Chinese-language scholarly publications related to the development and application of AI. This includes journal articles, conference papers, repository publications (such as arXiv), books, and theses.

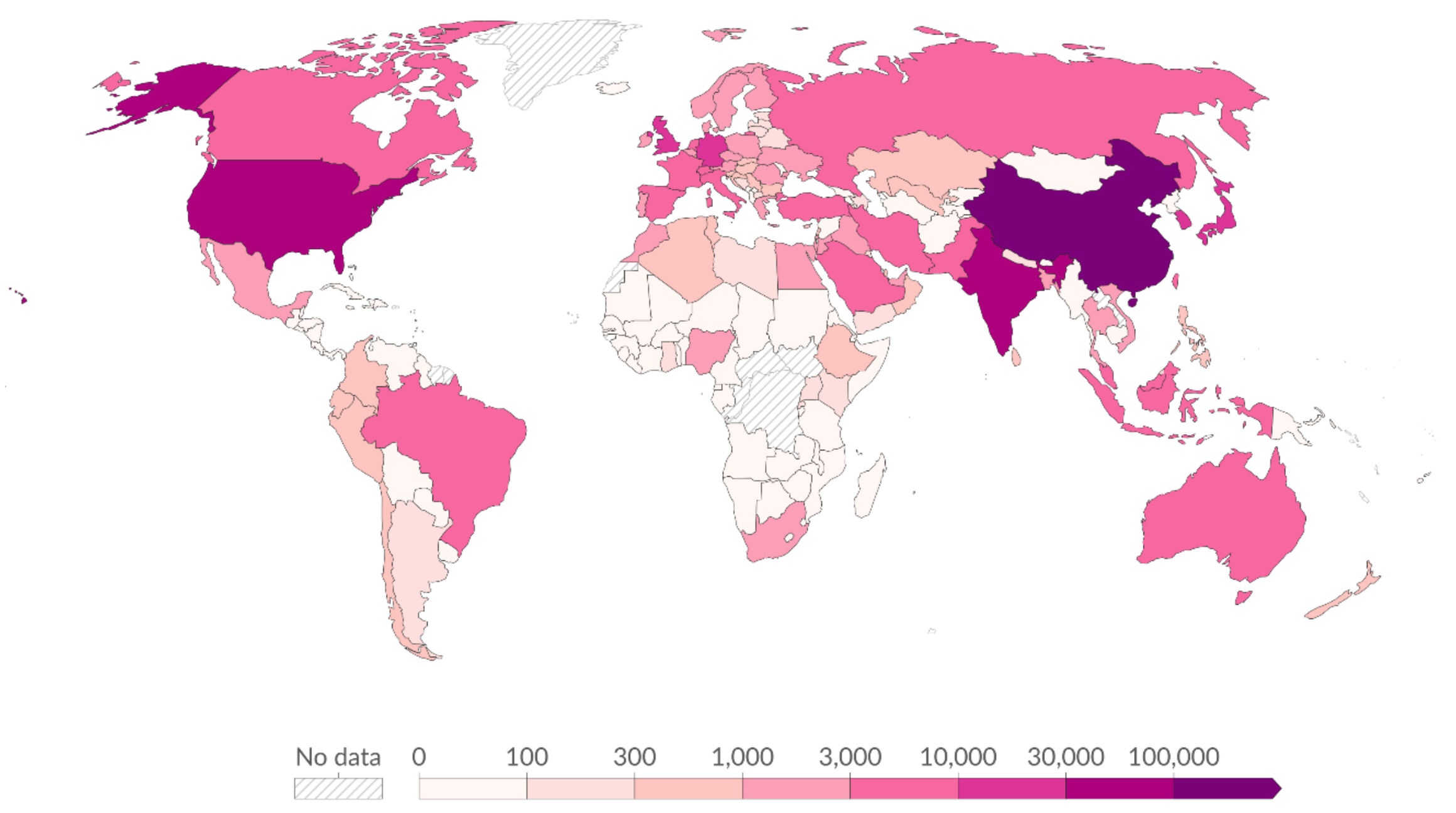

**Data source:** Center for Security and Emerging Technology (2025)

OurWorldinData.org/artificial-intelligence | 

*Figure 6: AI Scholarly Publications by Country for 2023 from Our World in Data*

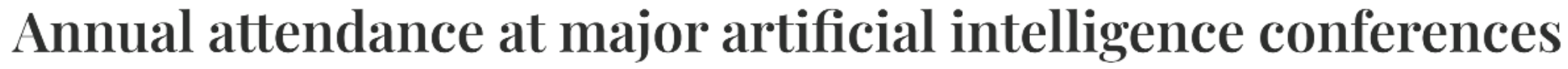

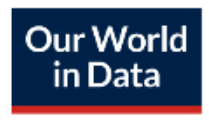

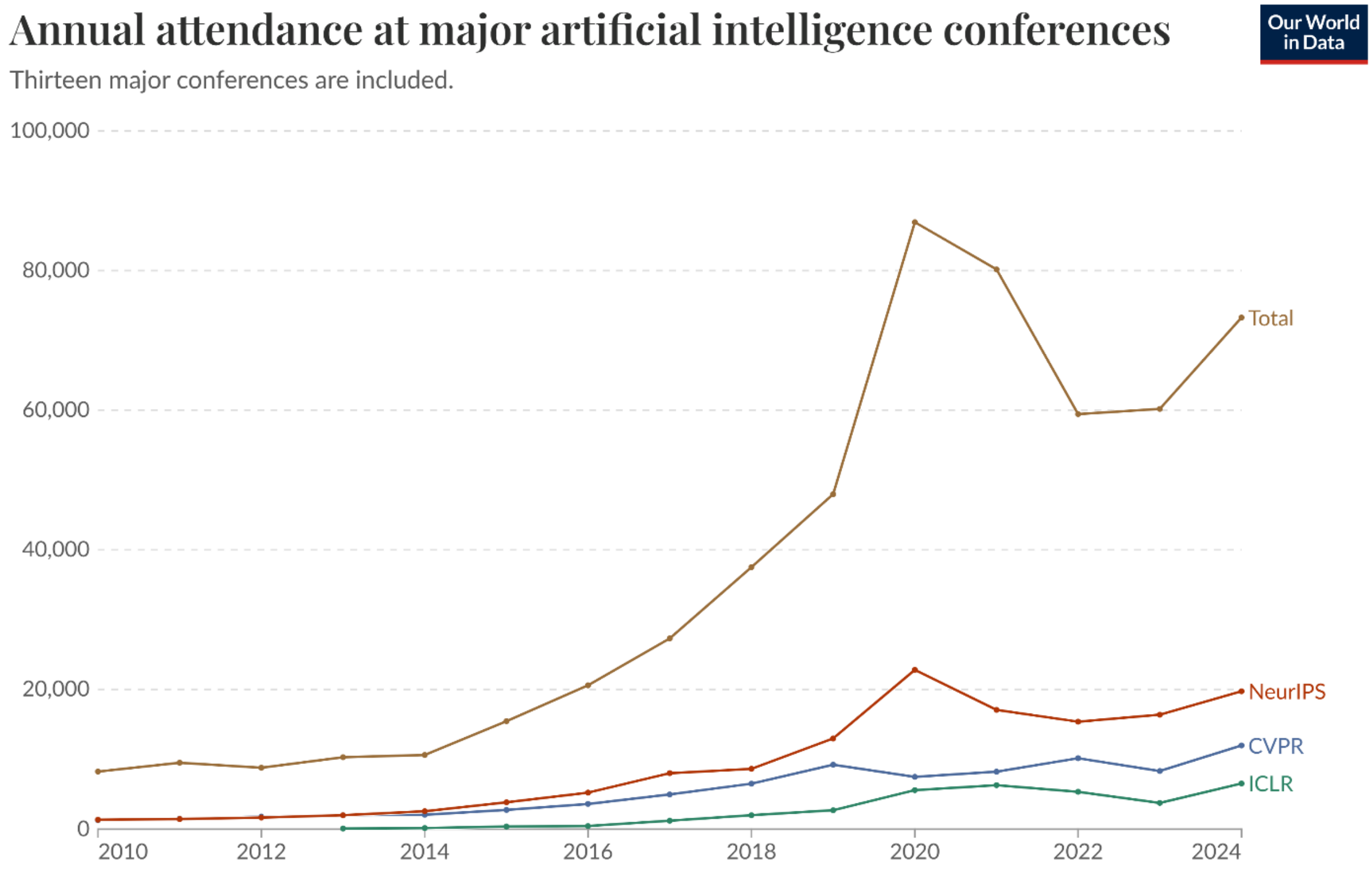

*Figure 7: Annual attendance at major AI conferences from Our World in Data*

## 2026 Baseline for Pacing Measures Growth

We set the baseline in 2026 for our pacing measure using averages of Epoch AI's Notable AI Model dataset [3]. We want to understand how growth in training compute (FLOP), training dataset size (FLOP), and training time (hours) are changing as well as the number of models within each generation.  For this we take the min, mean, median and max for each 2023-2025 for each pacing measure scaling hypothesis component in Table 4, Table 5, and Table 6 below.

*Table 4: Summary of Training Compute (FLOP) in Notable AI Models 2023-2025*

| Year | MIN | MEAN | MEDIAN | MAX | Models in Sample (n=) |
|---|---|---|---|---|---|
| 2023 | 4.90E+16 | 2.07E+24 | 6.33E+22 | 5.00E+25 | 65 |
| 2024 | 5.85E+20 | 6.23E+24 | 3.03E+24 | 3.80E+25 | 48 |
| 2025 | 9.62E+21 | 4.04E+25 | 4.08E+24 | 5.00E+26 | 37 |

*Table 5: Summary of Training Dataset Size (FLOP) in Notable AI Models 2023-2025*

| Year | MIN | MEAN | MEDIAN | MAX | Models in Sample (n=) |
|---|---|---|---|---|---|
| 2023 | 6.90E+04 | 4.13E+12 | 7.68E+10 | 1.60E+14 | 63 |
| 2024 | 2.40E+09 | 9.53E+12 | 7.00E+12 | 1.14E+14 | 33 |
| 2025 | 1.44E+05 | 1.67E+13 | 1.48E+13 | 3.60E+13 | 29 |

*Table 6: Summary of Training Time (hours) in Notable AI Models 2023-2025*

| Year | MIN | MEAN | MEDIAN | MAX | Models in Sample (n=) |
|---|---|---|---|---|---|
| 2023 | 1 | 944 | 347 | 6,528 | 41 |
| 2024 | 14 | 897 | 612 | 2,500 | 12 |
| 2025 | 48 | 1,715 | 1,572 | 3,240 | 13 |

The findings of Table 4 - Table 6 show the wide range in values for any individual model, and how changes in technology, (or increasing secrecy and reluctance to publish) results in widely varying maximum and minimum values. However the Median values for training compute (FLOP), training dataset size (FLOP), and training time (hours) steadily increases. We therefore use the following median 2025 values as the baseline starting measure for our pacing growth.

*Table 7: 2025 Median Values to Set Baseline 2026 Values for Pacing Measure*

| Year | Training Compute (FLOP) | Training Dataset Size (FLOP) | Training Time (hours) |
|---|---|---|---|
| 2025 Median Values | 4.08E+24 | 1.48E+13 | 1,572 |

## Pacing Measure Growth Factors

Using our previous analysis of 2023-2025 on notable AI models and comparing median results allows us to calculate growth factor in Table 8.

*Table 8: Calculating Growth Factors from Median Results 2023-2025*

| Year | Median Training Compute (FLOP) | Year of Year Growth Factor | Median Data Set (FLOP) | Year of Year Growth Factor | Median Training Time (hours) | Year of Year Growth Factor |
|---|---|---|---|---|---|---|
| 2023 | 6.33E+22 | | 7.68E+10 | | 347 | |
| 2024 | 3.03E+24 | 47.83 | 7.00E+12 | 91.15 | 612 | 1.76 |
| 2025 | 4.08E+24 | 1.35 | 1.48E+13 | 2.11 | 1,572 | 2.57 |

As Table 8 indicates the ushering in of LLM and their rapid evolution from 2023-2024 resulted in extremely high growth factors of x48 to x91 between Training Compute and Data Set Size. We therefore use the 2024-2025 growth factors in our comparison of year over year growth below.

*Table 9: Year over Year Growth Factors Estimated*

| Source | Growth Factor of Training Compute (FLOP) | Growth Factor of Data Set Size (FLOP) | Growth Factor in Training Time (hours) |
|---|---|---|---|
| Notable AI 2023-2025 Analysis | x1.35/year | x2.11 | x2.57 |
| Hardware Acquisition Cost Analysis**[15, p. 5]** | x2.4-3.0 with a doubling time of 9 months | | |
| | | | |

**Frontier Foundation Models Needed to Innovate a New Generation of Capability**

In our conceptual analogy, the evolution of one generation of military jet aircraft to the next is often best understood in retrospect or extensive analysis on well known capabilities. Given that we are just on the threshold of Agentic AI, it is difficult if not impossible to estimate what is required to 'innovate' a subsequent generation of Agentic AI capability. A good proxy might be the number of frontier foundation models, as these models often represent the cutting edge of capabilities. But there are few well researched estimates of how many frontier models must be innovated to achieve a next generation capability. We arbitrarily place this at five, and may revise in future editions as data.

# Bibliography

[1] G. Hamm, "2024 United States Data Center Energy Usage Report," Lawrence Berkely National Laboratory, 2025. doi: 10.71468/P1WC7Q.

[2] "Frontier Data Centers," Epoch AI. [Online]. Available: https://epoch.ai/data/data-centers

[3] D. Owen *et al.*, "Notable AI Models Dataset." Feb. 12, 2026. Accessed: Feb. 14, 2026. [Online]. Available: https://epoch.ai/data/ai-models-documentation#downloads

[4] A. Laurent, "NVIDIA HGX Platform: Data Center Physical Requirements Guide," Jan. 2026. [Online]. Available: https://intuitionlabs.ai/articles/nvidia-hgx-data-center-requirements

[5] S. Cruzes, "DATA CENTERS IN THE AGE OF AI: A TUTORIAL SURVEY ON INFRASTRUCTURE, SUSTAINABILITY, AND EMERGING CHALLENGES," Oct. 27, 2025, *Preprints*. doi: 10.36227/techrxiv.176158592.23065552/v1.

[6] "Glossary of Data Center Terms." [Online]. Available: https://www.pducables.com/resources/glossary-of-data-center-terms

[7] "NVIDIA DGX SuperPOD: Data Center Design Featuring NVIDIA DGX H100 Systems." [Online]. Available: https://docs.nvidia.com/dgx-superpod/design-guides/dgx-superpod-data-center-design-h100/latest/index.html#dgx-superpod-data-center-design-featuring-dgx-h100

[8] "How many servers does a data center have?," RackSolutions. [Online]. Available: https://www.racksolutions.com/news/blog/how-many-servers-does-a-data-center-have
[9] J. C. Hick and M. Izzo, "Futureproofing through 2035 for the AI and HPC Power Density Trend," Los Alamos National Labratory, Los Alamos, Oct. 2025.
[10] A. Katal, S. Dahiya, and T. Choudhury, "Energy efficiency in cloud computing data centers: a survey on software technologies," *Cluster Comput*, vol. 26, no. 3, pp. 1845–1875, Jun. 2023, doi: 10.1007/s10586-022-03713-0.
[11] D. Mytton, "Data centre water consumption," *npj Clean Water*, vol. 4, no. 1, p. 11, Feb. 2021, doi: 10.1038/s41545-021-00101-w.
[12] E. Keski-Nisula, "Demand response potential of AI data center facilities — Physical flexibility of cooling, energy storage, and backup power in Finnish wholesale and reserve electricity markets," Aalto University School of Electrical Engineering, 2025.
[13] A. Walters, A. Walker, and M. Kelley, "Rethinking AI Demand Part 1: AI Data Centers Are Experiencing a Surge of Training Demand - What Happens When the Surge is Over?" [Online]. Available: https://www.alvarezandmarsal.com/insights/rethinking-ai-demand-part-1-ai-data-centers-are-experiencing-surge-training-demand-what
[14] B. Srivathsan, M. Sorel, and P. Sachdeva, "AI power: Expanding data center capacity to meet growing demand," McKinsey. [Online]. Available: https://www.mckinsey.com/industries/technology-media-and-telecommunications/our-insights/ai-power-expanding-data-center-capacity-to-meet-growing-demand#/
[15] B. Cottier, R. Rahman, L. Fattorini, N. Maslej, T. Besiroglu, and D. Owen, "The rising costs of training frontier AI models," 2024, *arXiv*. doi: 10.48550/ARXIV.2405.21015.
[16] J. Sevilla *et al.*, "Can AI scaling continue through 2030?," Epoch AI. [Online]. Available: https://epoch.ai/blog/can-ai-scaling-continue-through-2030
[17] A. Souly *et al.*, "Poisoning Attacks on LLMs Require a Near-constant Number of Poison Samples," 2025, *arXiv*. doi: 10.48550/ARXIV.2510.07192.
[18] "Our World in Data - Artificial Intelligence," Our World in Data.